\documentclass[12pt]{article}

\usepackage{amsmath}
\usepackage{graphicx}
\usepackage{amsfonts}
\usepackage{amssymb}
\usepackage{latexsym}
\usepackage{color}
\usepackage{cite}
\input{colordvi.tex}

\renewcommand{\thefootnote}{\#\arabic{footnote}}

\newcommand{\rhor}{\rho_{r}}
\newcommand{\zetar}{\zeta_{r}}
\begin{document}
\setcounter{footnote}{0}

\begin{titlepage}
\begin{flushright}
RESCEU-21/11
\end{flushright}
\begin{center}


\vskip .5in

{\Large \bf
Temporal enhancement of super-horizon curvature perturbations from decays of two curvatons and its cosmological consequences
}

\vskip .45in

{\large
Teruaki Suyama$^1$
and 
Jun'ichi Yokoyama$^{1,2}$
}

\vskip .45in

{\em
$^1$
  Research Center for the Early Universe (RESCEU), Graduate School
  of Science,\\ The University of Tokyo, Tokyo 113-0033, Japan
  }\\
{\em
$^2$
  Institute for the Physics and Mathematics of the Universe (IPMU),\\
  The University of Tokyo, Kashiwa, Chiba, 277-8568, Japan
  }

\end{center}

\vskip .4in

\begin{abstract}
If more than one curvaton dominate the Universe at different epochs from each other,
 curvature perturbations can be temporarily enhanced to a value much larger than
the observed one $10^{-5}$. 
The traces of the enhancement may be left as higher order correlation functions,
that is, as non-Gaussianity, the stochastic gravitational waves that are sourced
by scalar-scalar mode couplings, as well as the primordial black holes that are formed
by the gravitational collapse of the enhanced curvature perturbations.
We first confirm that such a temporal enhancement indeed occurs by solving the linearized
perturbation equations both numerically and analytically.
We then derive an analytic expression of the full-order curvature perturbation 
which does not rely on the frequently used sudden decay approximation and is exact on super-horizon scales.
By using this analytic formula, we provide expressions of the non-linearity parameters 
$f_{\rm NL},~\tau_{\rm NL}$ and $g_{\rm NL}$.
If both two curvatons contribute to the final curvature perturbations,
the strong non-Gaussianity appears in the trispectrum rather than in the bispectrum.
We also find a unique consistency relation between $\tau_{\rm NL}$ and $g_{\rm NL}$ without $f_{\rm NL}$.
By using the second-order perturbation theory, we numerically show that
the spectrum of the induced gravitational waves has a plateau corresponding to
duration of the enhancement and such gravitational
waves can be probed by ultimate-DECIGO and space-based atomic interferometers. 
We finally calculate the abundance of the primordial black holes and put a constraint
on the amplitude of the enhanced curvature perturbations.
\end{abstract}
\end{titlepage}

\renewcommand{\thepage}{\arabic{page}}
\setcounter{page}{1}
\renewcommand{\thefootnote}{\#\arabic{footnote}}

\section{Introduction}
Cosmic observations \cite{Komatsu:2010fb} are now strongly supporting
the idea of primordial inflation \cite{i1,i2,i3}.
According to the inflationary scenario, the primordial perturbations, which are seeds for the inhomogeneous structure
of our Universe, are created from the field fluctuations that are generated quantum mechanically during inflation \cite{yu0,yu1,yu2,yu3}.
In the simplest inflationary scenario, the fluctuations of the inflaton, where
energy drives accelerated expansion, turn into the curvature perturbations when its
wavelength becomes larger than the Hubble length during inflation. 
Once generated, the curvature perturbation remains constant and this is the primordial perturbation that can be
compared with CMB observations.

This simple scenario, however, may not be the one that actually happened in our Universe.
For example, in the curvaton scenario \cite{Enqvist:2001zp,Lyth:2001nq,Moroi:2001ct}, a curvaton, a scalar field other than the inflaton, plays a role of
creating the primordial perturbation.
In this scenario, the curvaton perturbations turn into the curvature perturbation after inflation when
the curvaton decays into the radiation.
There are many other models in which the primordial perturbations are created after inflation such as inhomogeneous end of inflation \cite{Bernardeau:2002jf,Bernardeau:2004zz,Lyth:2005qk,Salem:2005nd,Alabidi:2006wa}, 
modulated reheating scenario \cite{Dvali:2003em,Kofman:2003nx} etc.. 
A common feature among all these models is that the curvature perturbation,
once generated from zero remains constant outside the horizon.

Although these nonstandard scenarios are already contrived, the real Universe may have evolved
even more complicated manner, say, with multiple curvatons dominating at different epochs.
Such a scenario has been studied in \cite{Choi:2007fya} and \cite{Assadullahi:2007uw},
the former focusing on the power spectrum and the latter on the bispectrum.
In \cite{Assadullahi:2007uw}, although not emphasized, it was implicitly shown that if the two curvatons dominate the Universe
at different epochs from each other, the curvature fluctuations may evolve in a dramatically different way
than the standard case.
That is, they can grow to an amplitude much larger than the observed value, $10^{-5}$,
when the first curvaton dominates and decays, and then they are moderated to the observed amplitude
when the second curvaton dominates and decays. Thus in this scenario the curvature perturbation
can be temporarily enhanced.

At first glance, this temporal enhancement seems to have little effect
on observables since it must occur, if at all, much
before the big-bang nucleosynthesis, not to mention the observation time. 
But this is not the case. We can provide at least three possible interesting consequences from this effect.

The first possibility is that, as pointed out in \cite{Assadullahi:2007uw},
large possibly detectable non-Gaussian perturbations can be generated.
If we denote by $\zeta_{\rm max}$ the maximum amplitude of the curvature perturbation when it is enhanced,
the so-called $f_{\rm NL}$ parameter is given by $f_{\rm NL}\simeq \zeta_{\rm max}/10^{-5} \gg 1$, under the
assumption that the curvature perturbation is sourced only by the first decaying curvaton fluctuations.
For example, if $\zeta_{\rm max}=10^{-3}$, then we get $f_{\rm NL}\simeq 100$.

The second possibility is generation of stochastic gravitational waves(GWs) whose peak frequency can fall into
a range of the GW detectors.
It is well known that at second order in perturbation the scalar-scalar coupling can source the GWs,
most efficiently when the scalar mode reenters the horizon \cite{Ananda:2006af,Baumann:2007zm,Saito:2008jc}.
Since the amplitude of GWs is proportional to the square of the curvature perturbations,
we expect that large amplitude of GWs can be generated at the horizon scales when the curvature perturbation
is being enhanced.

The third possibility is a formation of primordial black holes (PBHs) \cite{Hawking:1971ei,zel,Carr:1974nx}.
If $\zeta_{\rm max}$ is very large, the perturbation mode which reenters the horizon when the curvature
perturbation is being enhanced may undergo a gravitational collapse to form a black hole
\footnote{Some papers, such as \cite{Yokoyama:1995ex,Yokoyama:1998pt,Kawasaki:1997ju,Kawasaki:1998vx,Yokoyama:1999xi,Kawasaki:2007zz,Kawaguchi:2007fz}, also considered 
the generation of large amplitude of the curvature perturbations on particular scales 
and the PBH formation. Contrary to the two curvaton model considered in this paper,
the curvature perturbations in those models do not show the temporal enhancement
on super-horizon scales.}.
Since the abundance of PBHs on various masses are tightly constrained
from cosmic observations (see  \cite{Carr:2009jm} for the latest results),
we can limit a range of the parameters of the two curvaton model by using such constraints.

The present paper aims to take up this temporal enhancement of the curvature perturbation in two curvaton model, 
to provide a detailed analysis of the generation and evolution of fluctuations and to discuss three observational
implications mentioned above.

\section{Basic picture}
\label{sec.basic}
In this paper, we are interested in a situation where both curvatons dominate the Universe at different epochs.
Before going to the detailed analysis of the scenario, let us first explain four non-trivial assumptions needed for 
the designing scenario to work, along with the basic history of the universe under consideration.

The first assumption we will take in this paper is that, other than the inflaton, there are two light free scalar fields
(we call them $\sigma_1$-field and $\sigma_2$-field, respectively.) having VEVs smaller than the Planck scale in the early universe.
By light, we mean that masses of both curvatons are much smaller than the expansion rate of the Universe 
when it is reheated by the decay of the inflaton,
$m_1, m_2 \ll H_{\rm reh}$ with $m_1$ and $m_2$ being masses of the two curvatons and $H_{\rm reh}$ 
the Hubble parameter at the time of reheating.
This condition means that both curvatons are almost massless during inflation.
It is well known that such a light scalar field aquires classical fluctuation of order $H_{\rm inf}/2\pi$
on super-horizon scales.
The requirement of VEVs smaller than the Planck scale is to avoid the second inflation caused by a curvaton.
Although this requirement is not essential, we take it for simplicity.

After inflation ends, the inflaton oscillates around the minimum of the potential and
finally decays into radiation to complete reheating.
Then the radiation dominated universe starts.
At this stage, both curvatons are still subdominant. Therefore, those fluctuations have not
yet contributed to the curvature perturbation.

The next assumption is that both curvatons start to ocsillate during this radiation dominated epoch.
If the secondly decaying curvaton starts to oscillate after the first decaying curvaton decays,
the temporal enhancement of the curvature perturbation, which is the main focus of this paper, does not occur.
Therefore, we do not consider such a case.

Since the ocsillating free scalar field can be treated as non-relativistic particles,
we can assume that the universe consists of the radiation coming from inflaton, two non-relativistic particles that are not interacting.
During this epoch, the fraction of energy densities of both curvatons grows in proportion to the scale factor
while the ratio between energy densities of the two curvatons stays constant.
We call this era epoch A.

Without a loss of generality, we can assume that $\sigma_1$-field decays first and $\sigma_2$-field decays later, $\Gamma_1 \gg \Gamma_2$.
Then, the third assumption is that the energy density of $\sigma_1$-field in epoch A is much larger than that of $\sigma_2$-field
and there is a period when $\sigma_1$-field dominates the universe. We call such a period epoch B.
Since the $\sigma_1$-field is dominating the universe, it is this epoch when the $\sigma_1$-field perturbation is mostly
converted to the curvature perturbation.
By exactly the same reasoning as the single curvaton case, the curvature perturbation at this epoch is given by $\simeq \delta_1 \equiv \delta \rho_1/\rho_1$
(precise definition of the curvature perturbation and regorous calculations will be given later.).
Since the $\sigma_2$-field is subdominant during this epoch, keeping its energy fraction constant,
the $\sigma_2$-field fluctuations have not been converted to the curvature perturbations yet.
When the Hubble parameter becomes $\Gamma_1$, the $\sigma_1$-field decays into the radition and the universe is again dominated by the radiation.
We call a period dominated by such radiation epoch C.

Our last assumption is that the $\sigma_2$-field finally dominates the universe before it decays.
In other words, $\Gamma_2$ should be small enough to allow the $\sigma_2$-field to dominate the universe.
We call a period dominated by the $\sigma_2$-field epoch D.
During this epoch, since the universe evolves like a matter dominated universe, the radiation generated
from the decay of the $\sigma_1$-field is diluted.
Because of this dilution, the curvature perturbation coming from the radiation perturbation is reduced by a factor
$\Omega_r \ll 1$,
where $\Omega_r$ is a fraction of the radiation energy density to the total one.
Meanwhile, the $\sigma_2$-field contributes to the curvature perturbation by $\delta_2$.
When the Hubble parameter becomes equal to $\Gamma_2$, the $\sigma_2$-field decays into radiation and
the Universe is again dominated by the radiation.
Since there are no isocurvature perturbations any more, the curvature perturbation remains constant and
this should be regarded as the final perturbation that can be compared with the cosmological observations such as CMB.
Thus, the primordial perturbation at the linear order is estimated as 
\begin{equation}
{\rm final \,perturbation} \simeq \Omega_r \delta_1+\delta_2, \label{basicpicture}
\end{equation}
where $\Omega_r$ must be evaluated at the time when $\sigma_2$-field decays. 
From observations, we know that this is about $10^{-5}$.

From those arguments, we find that the curvature perturbation evolves from zero to $\delta_1$ at epoch B 
and then decays to $\Omega_r \delta_1+\delta_2$.
Just for an illustration, let us choose $\delta_1=10^{-2}, \Omega_r=10^{-4}$ and $\delta_2=10^{-5}$.
In this case, curvature perturbation is temporarily amplified to $10^{-2}$ and then decays to 
the observed value $10^{-5}$.

It is worthwhile to mention here that the existence of more than a single curvaton is essential
to make the curvature perturbations decay in time. 
One may consider a simpler situation where the inflaton generates large amplitude of the curvature perturbations
by the standard mechanism and a curvaton that acquires little quantum fluctuations during inflation 
dominates the Universe at late time. 
In this case, attenuation of the curvature perturbation due to the curvaton dominance does not happen.
Instead, the large amplitude of the curvature perturbations of the inflaton origin is taken over by the curvaton 
because the epoch  when the curvaton starts its oscillations is modulated.
We need at least two fields other than the inflaton to have a sensible model
in which the temporal enhancement happens.

In the following sections, we will give more quantitative discussions of this scenario and some interesting consequences.

\section{Decays of two curvatons and generation of perturbations}
\subsection{Evolution of the background quantities}
In this subsection, we see how the background quantities evolve in time.
The periods we focus on are from epochs A to D whose definitions are given in sec.\ref{sec.basic}.

The background spacetime is the spatially flat Friedmann-Lema\^\i
tre-Robertson-Walker (FLRW) universe whose metric is given by
\begin{equation}
ds^2=a^2 (\eta) \left( -d\eta^2+ \delta_{ij} dx^i dx^j \right)
= -dt^2+a^2(t) \delta_{ij} dx^i dx^j .
\end{equation}
Here $\eta$ is the conformal time.
As we mentioned in the last subsection, we can treat the dynamics of the curvatons as collections of
non-relativistic particles.
Therefore, the energy-momentum tensor for each curvaton can be written as
\begin{equation}
T^{\mu \nu}_Z=\rho_Z u^\mu_Z u^\nu_Z,
\end{equation}
where $Z$ runs $1$ and $2$ and $u^\mu_Z$ is the four-velocity of $Z$-field with a normalization condition $g_{\mu \nu}u^\mu_Z u^\nu_Z=-1$.

While the total energy-momentum tensor obeys the conservation law,
each energy-momentum tensor is no longer conserved because curvaton fields decay into radiation \cite{Kodama:1985bj,Malik:2002jb};
\begin{equation}
\nabla_\mu T^\mu_{(A)\nu}=Q_{(A)\nu},
\end{equation}
where $A$ now represents radiation, $1$ or $2$.
The r.h.s. represents the transfer of the energy and momentum of the fluid.

At the background order, $Q_{(a) \nu}$ for each fluid is given by
\begin{align}
&Q_{(1)0}=a\Gamma_1 \rho_1, \\
&Q_{(2)0}=a\Gamma_2 \rho_2, \\
&Q_{(r) 0}=-a\Gamma_1 \rho_1-a\Gamma_2 \rho_2.
\end{align}
Because of the isotropy of the background spacetime, all the spatial components are zero.

Therefore, the evolution equations for the background quantities are given by
\begin{align}
&{ \rho'_1}+3{\cal H} \rho_1= -a\Gamma_1 \rho_1, & 
{\dot \rho_1}+3{ H} \rho_1= -\Gamma_1 \rho_1,\label{backa}\\
&{\rho'_2}+3{\cal H} \rho_2= -a\Gamma_2 \rho_2,
&{\dot \rho_2}+3{ H} \rho_2= -\Gamma_2 \rho_2, \label{backb}\\
&{ \rho'_r}+4{\cal H} \rho_r= a\Gamma_1 \rho_1+ a\Gamma_2 \rho_2,
&{\dot \rho_r}+4{ H} \rho_r= \Gamma_1 \rho_1+ \Gamma_2 \rho_2, \label{backr}\\
&{\cal H}^2= \frac{8\pi G}{3} (\rho_1+\rho_2+\rho_r)a^2,
&{ H}^2= \frac{8\pi G}{3} (\rho_1+\rho_2+\rho_r),\label{backhubble}
\end{align}
where a prime and an overdot denote differentiation with respect to 
$\eta$ and $t$, respectively, with ${\cal H} \equiv { a'}/a$ and $H\equiv
\dot{a}/a$.
From the first two equations, we have
\begin{align}
&\rho_1(t)=\rho_{1,*}\left(\frac{a_* }{a(t)}\right)^3
e^{ -\Gamma_1 (t-t_*)},
\label{rhoone}\\
&\rho_2 (t)=\rho_{2,*}\left(\frac{a_* }{a(t)}\right)^3
e^{ -\Gamma_2 (t-t_*)},
\label{rhotwo}
\end{align}
where $t_*$ is an arbitrary time in the epoch A. 
Substituting these solutions into the third equation, we get
\begin{equation}
\rhor(t)=\rho_{r,*}\left(\frac{a_*}{a(t)}\right)^4
+\Gamma_1\int_{t_*}^t\left(\frac{a(t')}{a(t)}\right)^4\rho_1(t')dt'
+\Gamma_2\int_{t_*}^t\left(\frac{a(t')}{a(t)}\right)^4\rho_2(t')dt'. \label{rhor}
\end{equation}

\begin{figure}[t]
  \begin{center}{
    \includegraphics{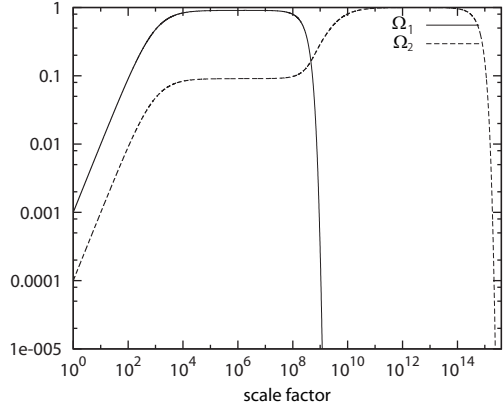}
    }
  \end{center}
  \caption{The evolution of $\Omega_1$ and $\Omega_2$. We chose $\Omega_{1,{\rm ini}}/\Omega_{2,{\rm ini}}=10$ and $\Gamma_2/\Gamma_1=10^{-10}$.}
 \label{fig:background}
\end{figure}

The designing situation in this paper is that both two curvatons dominate the universe at different epochs.
We give in Fig.\ref{fig:background} a typical evolution of $\Omega_1$ and $\Omega_2$.
We see that their evolution can be clearly devided into four epochs explained in the last section.
During epoch A, both of them grows as $\propto a$.
During epoch B, $\Omega_1 \simeq 1$ and $\Omega_2 \ll 1$ stays constant.
During epoch C, $\Omega_2$ starts to grow again like $\propto a$.
During epoch D, $\Omega_2 \simeq 1$ and this epoch terminates by the decay of the $\sigma_2$-field.  

\subsection{Linear order perturbation equations}
Before discussing perturbation behaviors and their consequences,
we give basic evolution equations for the scalar perturbations.

We write the perturbed metric of the scalar type in the conformal Newtonian gauge:
\begin{equation}
ds^2=a^2 (\eta) \big\{ -(1+2\psi)d\eta^2+(1-2\phi)\delta_{ij} dx^i dx^j \big\}. \label{scalarper}
\end{equation}
For the matter perturbations, we intoduce density contrast and the velocity perturbation by
\begin{equation}
\delta_A=\frac{\delta \rho_A}{\rho_A}, \hspace{5mm} u_{A i} = av^i_A.
\end{equation}
Note that $u_0=-a (1+\psi)$ is completely determined by the metric perturbation by using the
normalization condition.

At the linear order in the perturbation, $Q_{(A) \nu}$ for each fluid is given by \cite{Ichiki:2004vi}
\begin{align}
&\delta Q_{(1)0}=a \Gamma_1 \delta \rho_1+a \Gamma_1 \rho_1 \psi, \label{Qa-pert} \\
&\delta Q_{(2)0}=a \Gamma_1 \delta \rho_2+a \Gamma_2 \rho_2 \psi, \label{Qb-pert} \\
&\delta Q_{(r)0}=-a \Gamma_1 \delta \rho_1-a \Gamma_1 \rho_1 \psi-a \Gamma_2 \delta \rho_2-a \Gamma_2 \rho_2 \psi, \\
&\delta Q_{(1)i}=-a \Gamma_1 \rho_1 v^i_1, \\
&\delta Q_{(2)i}=-a \Gamma_2 \rho_2 v^i_2, \\
&\delta Q_{(r)i}=a \Gamma_1 \rho_1 v^i_1+a \Gamma_2 \rho_2 v^i_2. 
\end{align}
The second terms in (\ref{Qa-pert}) and (\ref{Qb-pert}) represent an effect due to the modulation of time. 
$Q_{(r) \mu}$ for the radiation is obtained from the conservation law for the total energy momentum.
Using these equations, we find that the continuity equation and the Euler equation for each fluid are given by
\begin{align}
&{\delta'_1}-3 { \phi'}+k v_1 =-a \Gamma_1 \psi, \\
&{ \delta'_2}-3 {\phi'}+k v_2 =-a \Gamma_2 \psi, \\
&{ \delta'_r}-4 {\phi'}+\frac{4}{3} k v_r=a \Gamma_1 \frac{\rho_1}{\rho_r} \left( \delta_1-\delta_r+ \psi \right)+a \Gamma_2 \frac{\rho_2}{\rho_r} \left( \delta_2-\delta_r+ \psi \right), 
\end{align}
and
\begin{align}
&{ v'_1}+{\cal H} v_1-k \psi=0, \\
&{ v'_2}+{\cal H} v_2-k \psi=0, \\
&{ v'_r}-\frac{k}{4} \delta_r-k\psi=a \Gamma_1 \frac{\rho_1}{\rho_r} \left( \frac{3}{4}v_1-v_r \right)+a \Gamma_2 \frac{\rho_2}{\rho_r} \left( \frac{3}{4}v_2-v_r \right),
\end{align}
respectively, where $k$ is the comoving wavenumber.
From the perturbed Einstein equations, we can derive the following evolution equations:
\begin{align}
&k^2 \phi+3{\cal H} { \phi'}+3{\cal H}^2 \psi=-4\pi G (\rho_1 \delta_1+\rho_2 \delta_2+\rho_r \delta_r)a^2, \label{ein00}\\
&k ({\cal H} \psi+{ \phi'})=4\pi G a^2 \left( \rho_1 v_1+\rho_2 v_2+\frac{4}{3} \rho_r v_r \right), \\
&\phi-\psi=0. \label{eintraceless}
\end{align}

\subsection{Perturbation evolution on super-horizon scales}
Since understanding the evolution of the curvature perturbation on super-horizon scales is important for our purposes,
let us next consider the perturbation behavior on super-horizon scales,
postponing the analysis on sub-horizon scales of perturbations.
The evolution equations on super-horizon scales can be obtained by setting $k=0$ in the perturbation
equations we gave in the last subsection.

\begin{figure}[t]
  \begin{center}{
    \includegraphics{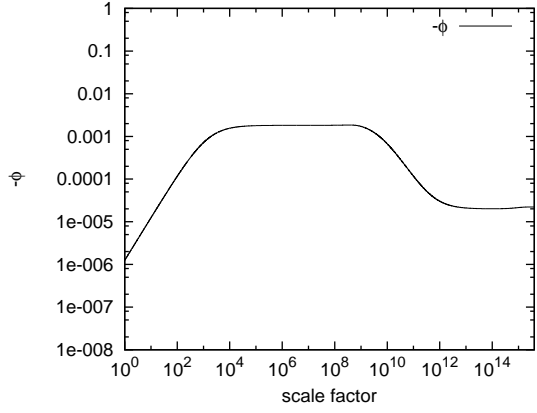}
    }
  \end{center}
  \caption{The evolution of $\phi$. Since $\phi$ is negative, we plot $-\phi$. 
We chose $\Omega_{1,{\rm ini}}/\Omega_{2,{\rm ini}}=10, \delta_{1,{\rm ini}}=10^{-2}, \delta_{2,{\rm ini}}=10^{-4}, \Gamma_2/\Gamma_1=10^{-10}$.}
 \label{fig:perturbation}
\end{figure}

In Fig.\ref{fig:perturbation}, we show a typical evolution of $\phi$ which is obtained by solving numerically the 
perturbation equations with $k=0$.
We started the calculation at a time when the  expansion rate $H$ 
is much larger than
$\Gamma_1$, with the initial conditions in the epoch A given by the analytic approximation:
\begin{align}
\delta_1(\eta)&=\delta_{1,{\rm ini}}-\frac{3}{8} \left( \delta_{1,{\rm ini}} \Omega_1(\eta)+\delta_{2,{\rm ini}} \Omega_2(\eta) \right) +\cdots, \label{inia}\\
\delta_2(\eta)&=\delta_{2,{\rm ini}}-\frac{3}{8} \left( \delta_{1,{\rm ini}} \Omega_1(\eta)+\delta_{2,{\rm ini}} \Omega_2(\eta) \right) +\cdots, \label{inib}\\
\delta_r(\eta)&=-\frac{1}{2}\left( \delta_{1,{\rm ini}} \Omega_1(\eta)+\delta_{2,{\rm ini}} \Omega_2(\eta) \right)+\cdots,\\
\phi(\eta)&=-\frac{1}{8}\left( \delta_{1,{\rm ini}} \Omega_1(\eta)+\delta_{2,{\rm ini}} \Omega_2(\eta) \right)+\cdots,
\end{align}
where $\cdots$ represent terms that are suppressed at early time with higher powers of $\Omega_1$ and/or
$\Omega_2$.
Since we assume that inflaton fluctuation contributes little to the curvature perturbation,
we have imposed the condition that both $\delta_r$ and $\phi$ vanish at the outset.
The parameters used in Fig.\ref{fig:perturbation} are such that
$\delta_{1,{\rm ini}}=10^{-2}$ and $\delta_{2,{\rm ini}}=10^{-4}$.
The parameters for the background are the same as the ones used in Fig.\ref{fig:background}.

From Fig.\ref{fig:background}, we can clearly see that $\phi$ is enhanced during epoch B.
The order of $\phi$ at this time is roughly $\delta_{1,{\rm ini}}$ (we will provide exact analytic expression later.).
This enhancement still persists during epoch C until the $\sigma_2$-field dominates the universe.
As the $\sigma_2$-field dominates, $\phi$ decays to $\Omega_{r,2} \delta_{1,{\rm ini}}+\delta_{2,{\rm ini}}$,
where $\Omega_{r,2} \ll 1$ is $\Omega_r$ evaluated at the time when the $\sigma_2$-field decays.
Therefore, the enhanced value of the curvature perturbation is determined by $\delta_{1,{\rm ini}}$
and the final value is determined by either $\Omega_{r,2} \delta_{1,{\rm ini}}$ or $\delta_{2,{\rm ini}}$,
whichever is greater.

We can analytically derive time evolution of curvature perturbations
in the super-horizon regime
solving  perturbation equations for gauge-invariant
variable,
$\zeta_A$, which is defined by $\zeta_A\equiv
-\phi-H\frac{\delta\rho_A}{\dot \rho_A}$ and represents curvature
perturbation
on the uniform energy density surface of the component $A$.  
In the presence of interactions their evolution equations in the 
super-horizon ($k\longrightarrow 0$) limit read
\begin{eqnarray}
\dot{\zeta_1}&=&-\frac{1}{3}\Gamma_1\zeta_1, \\
\dot{\zeta_2}&=&-\frac{1}{3}\Gamma_2\zeta_2, \\
\dot{\zeta_r}&=&\frac{\Gamma_1\rho_1}{4\rho_r}\left[
-(4+3\gamma_1)\zeta_r+(3+3\gamma_1)\zeta_1
\right]+\frac{\Gamma_2\rho_2}{4\rho_r}\left[
-(4+3\gamma_2)\zeta_r+(3+3\gamma_2)\zeta_2
\right], \label{zetareq}
\end{eqnarray}
wehre $\gamma_A\equiv \Gamma_A/(3H)$.  
Using the background solutions for $\rho_1(t)$, $\rho_2(t)$, and
$\rhor(t)$,
namely (\ref{rhoone}), (\ref{rhotwo}), and (\ref{rhor}), these equations
are solved as
\begin{eqnarray}
\zeta_1(t)&=&\zeta_{1*}e^{-\frac{1}{3}\Gamma_1(t-t_*)},\\
\zeta_2(t)&=&\zeta_{2*}e^{-\frac{1}{3}\Gamma_1(t-t_*)},\\
\zeta_r(t)&=&\int_{t_*}^t\left[
\frac{3\Gamma_1\rho_1(t')}{4\rhor(t')}(1+\gamma_1)\zeta_{1*}
e^{-\frac{1}{3}\Gamma_1(t'-t_*)}+
\frac{3\Gamma_2\rho_2(t')}{4\rhor(t')}(1+\gamma_2)\zeta_{2*}
e^{-\frac{1}{3}\Gamma_2(t'-t_*)}
\right] \nonumber\\
&&~~~\times
\exp\left\{-\int_{t'}^t\left[
\frac{\Gamma_1\rho_1(t'')}{4\rhor(t'')}(4+3\gamma_1)+
\frac{\Gamma_2\rho_2(t'')}{4\rhor(t'')}(4+3\gamma_2)
\right]dt''\right\}dt'.\label{zetasolr}
\end{eqnarray}
In terms of these solutions the 
curvature perturbation in the uniform total density surface,
$\zeta$, is expressed as
\begin{equation}
\zeta\equiv-\phi-H\frac{\delta\rho}{\dot\rho}
=\frac{3\rho_1\zeta_1+3\rho_2\zeta_2+4\rho_r\zeta_r+3\gamma_1\rho_1(\zeta_1-\zeta_r)+3\gamma_2\rho_2(\zeta_2-\zeta_r)}{{3\rho_1+3\rho_2+4\rho_r}}.
\end{equation}
The key quantity to understand the time evolution of $\zeta$ is
$\zeta_r$.
In the regimes A and B, only the terms involving $\rho_1$ is important
in the both integrands in (\ref{zetasolr}) because by assumption 
$\rho_2$ is much smaller
than $\rho_1$ then.  Furthermore, in the regime B when $\rho_1 \gg
\rhor$ the last exponential factor in the right-hand-side of
(\ref{zetasolr}) takes an appreciable value of order of unity only 
for a short time interval 
\begin{equation}
\Delta t''=t-t'\approx \frac{4\rhor}{\Gamma_1\rho_1(4+3\gamma_1)}, 
\end{equation}
which also limits the range of $t'$ integral.  As a result we find
\begin{equation}
\zeta_r(t) \approx \frac{3+3\gamma_1}{4+3\gamma_1}\zeta_1(t).
\end{equation}
We can estimate $\zetar$ at the end of the regime B by the 
contribution at the epoch when the ratio $\rho_1/\rhor$
is the largest.  This is just before $\sigma_1$'s decay when $\gamma_1$
was still negligible with $\zeta_1(t)\approx \zeta_{1*}$.
We therefore find $\zeta_r\approx 3\zeta_{1*}/4$ at the end of 
the regime B.

The behavior of $\zetar$ in the regime D can also be understood 
similarly replacing the suffix 1 by 2.  When $\rho_2$ dominates
over $\rhor$, we find
$\zetar(t)\approx 3\zeta_{1*}/4+3\zeta_{2*}/4$, so that the total
curvature
perturbation reads
\begin{equation}
\zeta(t)\approx
 \frac{3\rho_2(t)\zeta_{2*}+3\rhor(t)(\zeta_{1*}+\zeta_{*2})}
{3\rho_2(t)+4\rho_r(t)}, 
\end{equation}
which clearly shows that as the fraction $\rho_2(t)/\rhor(t)$ increases
the large curvature perturbation due to $\zeta_{1*}$ is regulated 
to a smaller value to reach the final value
\begin{equation}
\zeta\approx \left(1+\left.\frac{\rhor}{\rho_2}\right|_d\right)\zeta_{2*}
+\left.\frac{\rhor}{\rho_2}\right|_d\zeta_{1*}
\approx \zeta_2+\left(\frac{\Gamma_2}{\Gamma_1}\right)^{2/3}
\left(\frac{\Omega_{1,*}}{\Omega_{2,*}}\right)^{4/3}\zeta_{1*}.
\label{zz}
\end{equation}
Here $\left.\frac{\rhor}{\rho_2}\right|_d\equiv \left(\frac{\Gamma_2}{\Gamma_1}\right)^{2/3}\left(\frac{\Omega_{1,*}}{\Omega_{2,*}}\right)^{4/3}$ denotes the ratio of
radiation
energy from $\rho_1$ to $\rho_2$ upon decay of $\sigma_2$.
Note that $\zeta_{1,*}=\delta_{1,*}/3$  and $\zeta_{2,*}=\delta_{2,*}/3$ 
hold because $\phi$ 
is negligible and $\Gamma_1,~\Gamma_2 \ll H$ at the outset.
 
\subsection{Nonlinear super-horizon perturbations}
Eq.~(\ref{zz}) clarifies how the final curvature perturbation after its temporal enhancement 
is related to the initial amplitudes of the two curvatons.
Although an essential point is completely manifested by Eq.~(\ref{zz}), the approximations 
we have made to derive it neglects a ${\cal O}(1)$ constant factor in front of the second term in Eq.~(\ref{zz}).

In this subsection, by using $\delta N$ formalism \cite{Starobinsky:1986fxa,Sasaki:1995aw,Nambu:1997wh,Sasaki:1998ug,Lyth:2004gb},
we derive the exact analytic expressions for the enhanced and the final amplitudes
of the curvature perturbation not only to linear order in perturbation but also to
any higher order.
According to this formalism, the curvature perturbation on the uniform total energy density hypersurface
at a point ${\vec x}$ is given by the perturbation of the e-folding number:
\begin{align}
\zeta (\eta,{\vec x})&=N(\sigma_1(\eta,{\vec x}),\sigma_2(\eta,{\vec x}))-{\rm spatial \,average}, \nonumber \\
&=\int_{\eta_*}^\eta d\eta' \, {\cal H} (\eta',\sigma_1(\eta',{\vec x}),\sigma_2(\eta',{\vec x}))-{\rm spatial \,average}, \label{deltaN}
\end{align}
where $\eta_*$ is an arbitrary earlier time than $\eta$ and the hypersurface at $\eta_*$ should be
the flat slicing.
The evolutions of $\sigma_1,\sigma_2$ and ${\cal H}$ are determined by solving the background equations (\ref{backa})-(\ref{backhubble}).

In the appendix, we show that the number of e-folds from $\eta_*$ at the epoch A to $\eta_f$ well after the $\sigma_2$-field decay is given by
\begin{align}
N(\sigma_1(\eta_f,{\vec x}),\sigma_2(\eta_f,{\vec x}))=&\frac{1}{4} \log \bigg\{ \left( 1+\epsilon_\Gamma \right) {\left( \frac{{\cal H}_*}{a_* \Gamma_1} \right)}^{2/3} \Omega_{1,*}^{4/3} (\eta_*,{\vec x})+{\left( \frac{{\cal H}_*}{a_* \Gamma_2} \right)}^{2/3} \Omega_{2,*}^{4/3}(\eta_*,{\vec x}) \bigg\} \nonumber \\
&+\frac{1}{4}\log \bigg\{ {\left( \frac{9}{4} \right)}^{1/3} a_* c_\Gamma \bigg\} +\frac{1}{4} \log \frac{\rho_*}{\rho_f}, \label{full-efolding}
\end{align}
where $\epsilon_\Gamma \approx 1.183$ is a numerical value and $\rho_*/\rho_f$ is the total energy density
at $\eta_*/\eta_f$. $c_\Gamma$ is another numerical value.
Eq.\,(\ref{full-efolding}) is obtained without using the sudden decay approximation which is frequently adopted in the literature.
In the limiting case where both curvatons dominate the universe at different epochs separated far enough, (\ref{full-efolding}) is exact.
Therefore, by combining (\ref{full-efolding}) with (\ref{deltaN}), we can derive the exact expression of $\zeta$ which is correct 
to any order in the perturbation.
We can relate $\Omega_{1,*}(\eta_*,{\vec x})$ and $\Omega_{2,*}(\eta_*,{\vec x})$ with the density contrast for each field as
\begin{equation}
\Omega_{1,*} (\eta_*,{\vec x})=\Omega_{1,*}\left[ 1+\delta_1 (\eta_*,{\vec x}) \right], \hspace{5mm} \Omega_{2,*} (\eta_*,{\vec x})=\Omega_{2,*} \left[ 1+\delta_2 (\eta_*,{\vec x}) \right].
\end{equation}
If we choose $\eta_*$ to be deep in the epoch A, each density contrast can be approximated with
the initial density contrast (see Eqs.\,(\ref{inia}) and (\ref{inib})):
\begin{equation}
\Omega_{1,*} (\eta_*,{\vec x})=\Omega_{1,*}\left[ 1+\delta_{1,{\rm ini}} ({\vec x}) \right], \hspace{5mm} \Omega_{2,*} (\eta_*,{\vec x})=\Omega_{2,*} \left[ 1+\delta_{2,{\rm ini}} ({\vec x}) \right].
\end{equation}
The initial density contrast is determined by quantum fluctuations generated when the mode crossed
the Hubble length during inflation.

Using these equations, we find that the full non-linear expression of $\zeta$ is given by
\begin{align}
\zeta (\eta_f,{\vec x})=&\frac{1}{4} \log \bigg\{ (1+\epsilon_\Gamma) \Omega_1^{4/3}(\eta_*) {[1+\delta_{1,{\rm ini}} ({\vec x})]}^{4/3}+{\left( \frac{\Gamma_1}{\Gamma_2} \right)}^{2/3} \Omega_2^{4/3}(\eta_*){[1+\delta_{2,{\rm ini}} ({\vec x}) ]}^{4/3} \bigg\} \nonumber \\
&-\frac{1}{4} \log \bigg\{ (1+\epsilon_\Gamma) \Omega_1^{4/3}(\eta_*)+{\left( \frac{\Gamma_1}{\Gamma_2} \right)}^{2/3} \Omega_2^{4/3}(\eta_*) \bigg\}. \label{full-zeta}
\end{align}
At linear order in the density contrasts, this equation reduces to
\begin{align}
\zeta (\eta_f,{\vec x})&=\frac{1}{3} \frac{(1+\epsilon_\Gamma) {( \frac{\Gamma_2}{\Gamma_1} )}^{2/3} \Omega_{1,*}^{4/3} \delta_{1,{\rm ini}} ({\vec x})+\Omega_{2,*}^{4/3} \delta_{2,{\rm ini}} ({\vec x})}{(1+\epsilon_\Gamma) {( \frac{\Gamma_1}{\Gamma_2} )}^{2/3} \Omega_{1,*}^{4/3}+\Omega_{2,*}^{4/3}}, \nonumber \\
&\approx \frac{1}{3} (1+\epsilon_\Gamma) {\left( \frac{\Gamma_2}{\Gamma_1} \right)}^{2/3} {\left( \frac{\Omega_1}{\Omega_2} \right)}^{4/3} \delta_{1,{\rm ini}} ({\vec x})+\frac{1}{3} \delta_{2,{\rm ini}} ({\vec x}).
\end{align}
To get the second equation, we have used an inequality ${( \frac{\Gamma_2}{\Gamma_1} )}^{2/3} \Omega_{1,*}^{4/3} \ll \Omega_{2,*}^{4/3}$
which is equivalent to the condition that the $\sigma_2$-field dominates the Universe eventually.
Therefore, it is useful to introduce a parameter $s$ defined by
\begin{equation}
s \equiv (1+\epsilon_\Gamma){\left( \frac{\Gamma_2}{\Gamma_1} \right)}^{2/3} {\left( \frac{\Omega_{1,*}}{\Omega_{2,*}} \right)}^{4/3},
\end{equation}
which roughly represents the fraction of the radiation from $\sigma_1$-field decay at the time of $\sigma_2$-field decay.
With $s$, $\zeta$ becomes
\begin{equation}
\zeta (\eta_f,{\vec x}) \approx \frac{1}{3} s \delta_{1,{\rm ini}} ({\vec x})+\frac{1}{3} \delta_{2,{\rm ini}} ({\vec x}), \label{linearzeta}
\end{equation}
which is a more regorous expression of (\ref{zz}).
We see that the transfer coefficient of $\delta_1$ is proportional to $s$.
Apart from the numerical factors, Eq.~(\ref{linearzeta}) is nothing more than
the rough estimation (\ref{basicpicture}) and also agrees with the result of \cite{Assadullahi:2007uw}.

We can also derive the maximum magnitude of the enhanced $\zeta$ during the epochs B and C.
Since the $\sigma_2$-field is subdominant during those epochs, $\zeta_{\rm max}$ is completely sourced
by the $\sigma_1$-field perturbation.
This means that $\zeta_{\rm max}$ is equal to the final curvature perturbation in the single curvaton
model in which the curvaton dominates the universe before its decay.
Therefore, $\zeta_{\rm max}$ can be obtained by the second term in (\ref{linearzeta}) with $\delta_2$ 
replaced by $\delta_1$:
\begin{equation}
\zeta_{\rm max} ({\vec x}) = \frac{1}{3} \delta_{1,{\rm ini}} ({\vec x}).
\end{equation}

We can convert $\zeta(\eta_f)$ and $\zeta_{\rm max}$ into the corresponding $\phi(\eta_f)$ and $\phi_{\rm max}$.
In the linear perturbation theory, it is well known that $\phi$ is related to $\zeta$ by \cite{Liddle-Lyth}
\begin{align}
\phi&=-\frac{2}{3} \zeta, \hspace{5mm} {\rm (for\,the\,radiation\,dominated\,universe)}, \\
\phi&=-\frac{3}{5} \zeta, \hspace{5mm} {\rm (for\,the\,matter\,dominated\,universe)}.
\end{align}
Therefore, $\phi(\eta_f)$ becomes
\begin{equation}
\phi(\eta_f)=-\frac{2}{9} s \delta_{1,{\rm ini}}-\frac{2}{9} \delta_{2,{\rm ini}}.
\end{equation}
On the other hand, $\phi_{\rm max}$ at the epoch B/C is given by
\begin{align}
\phi_{\rm max}&=-\frac{1}{5} \delta_{1,{\rm ini}}, \hspace{5mm} {\rm (in\,the\,epoch\,B)}, \\
\phi_{\rm max}&=-\frac{2}{9} \delta_{1,{\rm ini}}, \hspace{5mm} {\rm (in\,the\,epoch\,C)}.
\end{align}

\section{Implications}
We have shown that the curvature perturbation can be temporarily enhanced in two curvaton models.
At first glance, it seems that such an enhancement has nothing to do with observations since what 
we observe is the curvature perturbation at or after the time of last
scattering epoch when it 
has already settled down to the observed value $10^{-5}$.
This naive guess is true for the power spectrum.
However, the trace of the enhancement enters the game when we consider the black hole formation,
higher order correlation functions of the curvature perturbation (non-Gaussianity),
and the gravitational waves generated by scalar-scalar mode couplings.

\subsection{Non-Gaussianity}
Eq.~(\ref{full-zeta}) is the fully non-linear expression of the curvature perturbation.
By using this equation, we can calculate correlation functions of any order.
In this paper, we calculate the three and four-point functions (bi- and tri-spectra) which
are now becoming important observables to extract information of the early universe.

The curvature perturbation given by Eq.~(\ref{full-zeta}) is the so-called local type
for which the curvature perturbation depends on the source fields at the same point \cite{Lyth:2005fi}
\begin{equation}
\zeta ({\vec x}) = N_a \delta \sigma_a({\vec x}) +\frac{1}{2} N_{ab} \delta \sigma_a ({\vec x})\delta \sigma_b({\vec x})+\frac{1}{6} N_{abc} \delta \sigma_a({\vec x}) \delta \sigma_b({\vec x}) \delta \sigma_c({\vec x})+\cdots,
\end{equation}
where $\delta \sigma_a ({\vec x})$ is the Gaussian field fluctuation at a point ${\vec x}$ at
some initial time and $N_a=\partial N/\partial\sigma_a$.

The power spectrum, $P_\zeta$, bispectrum, $B_\zeta$, and trispectrum,
$T_\zeta$, of the curvature perturbation are defined by
\begin{equation}
\label{eq:power}
\langle \zeta_{\vec k_1} \zeta_{\vec k_2} \rangle
=
{(2\pi)}^3 P_\zeta (k_1) \delta ({\vec k_1}+{\vec k_2}),
\end{equation}
\begin{eqnarray}
\langle \zeta_{\vec k_1} \zeta_{\vec k_2} \zeta_{\vec k_3} \rangle
&=&
{(2\pi)}^3 B_\zeta (k_1,k_2,k_3) \delta ({\vec k_1}+{\vec k_2}+{\vec k_3}),
\label{eq:bi}
\end{eqnarray}
and
\begin{eqnarray}
\langle
\zeta_{\vec k_1} \zeta_{\vec k_2} \zeta_{\vec k_3} \zeta_{\vec k_4}
\rangle
&=&
{(2\pi)}^3 T_\zeta (k_1,k_2,k_3,k_4) \delta ({\vec k_1}+{\vec k_2}+{\vec k_3}+{\vec k_4}),
\label{eq:tri}
\end{eqnarray}
respectively.
For the case of local type curvature perturbation, $B_\zeta$ and $T_\zeta$ can be written as
\begin{eqnarray}
B_\zeta (k_1,k_2,k_3)
&=&
\frac{6}{5} f_{\rm NL}
\left(
P_\zeta (k_1) P_\zeta (k_2)
+ P_\zeta (k_2) P_\zeta (k_3)
+ P_\zeta (k_3) P_\zeta (k_1)
\right), \\
\label{eq:def_f_NL}
T_\zeta (k_1,k_2,k_3,k_4)
&=&
\tau_{\rm NL} \left(
P_\zeta(k_{13}) P_\zeta (k_3) P_\zeta (k_4)+11~{\rm perms.}
\right) \nonumber \\
&&
+ \frac{54}{25} g_{\rm NL} \left( P_\zeta (k_2) P_\zeta (k_3) P_\zeta (k_4)
+3~{\rm perms.} \right),
\label{eq:def_tau_g_NL}
\end{eqnarray}
with $k_{13} = |{\vec k_1} + {\vec k_3}|$.  
Here the constant parameters $f_{\rm NL}, \tau_{\rm NL}$ and $g_{\rm NL}$ are the so-called
non-linearity parameters and given by \cite{Byrnes:2006vq}
\begin{align}
{6 \over 5}f_{\rm NL}&= \frac{N_a N_b N^{ab}}
{\left( N_c N^c \right)^2}, \label{eq:fNL} \\
\tau_{\rm NL}&= \frac{N_a N_{b} N^{ac} N_c^{~b}} 
{\left( N_d N^d \right)^3},
\label{eq:tauNL}
\end{align}
and
\begin{eqnarray}
\frac{54}{25}g_{\rm NL}=\frac{N_{abc} N^a N^b N^c}
{\left( N_d N^d \right)^3}.
\label{eq:gNL}
\end{eqnarray}

Let us define a new parameter $r$ by
\begin{equation}
r \equiv s\sigma_1/ \sigma_2 \approx \delta_{2,{\rm ini}}/(s\delta_{1,{\rm ini}}), \label{def-r}
\end{equation}
which represents the contribution of $\delta_{2,{\rm ini}}$ to $\zeta$ compared to that of
$\delta_{1,{\rm ini}}$.
If $r=0$, then $\zeta$ is solely sourced by the fluctuations in $\sigma_1$.
If $r \gg 1$, then $\zeta$ is mostly sourced by the fluctuations in $\sigma_2$.
With this parameter, the non-linearity parameters for two curvaton case are given by
\begin{align}
f_{\rm NL} &= -\frac{(15r^2+80)r^2s-25}{12{(r^2+1)}^2s}=\frac{25}{12 {(r^2+1)}^2s}+{\cal O}(1), \label{twofnl} \\
\tau_{\rm NL}&=\frac{(9r^4+112r^2+64)r^2s^2-80r^2s+25}{4 {(r^2+1)}^3s^2}=\frac{25}{4{(r^2+1)}^3s^2}+{\cal O} \left( \frac{1}{s} \right), \label{twotnl}\\
g_{\rm NL}&=\frac{(225r^4+3300r^2)r^2s^2-1500r^2s+125}{108 {(r^2+1)}^3s^2}=\frac{125}{108{(r^2+1)}^3s^2}+{\cal O} \left( \frac{1}{s} \right). \label{twognl}
\end{align}
Since $s$ enters $f_{\rm NL}$ in the denominator,
$s \ll 1$, which is satisfied in the situation we are interested in, 
is a necessary condition to have large $f_{\rm NL}$.
This condition can be qualitatively understood by expanding (\ref{linearzeta}) to second order in $\delta \sigma_1$:
\begin{equation}
\zeta= \frac{1}{3}s \left( 2\frac{\delta \sigma_1}{\sigma_1}+ {\left( \frac{\delta \sigma_1}{\sigma_1} \right)}^2 \right)=\zeta_g+\frac{3}{4s}\zeta_g^2, \label{quasi-second}
\end{equation}
where we have set $\delta_2=0$ and $\zeta_g \equiv \frac{2}{3}s \frac{\delta \sigma_1}{\sigma_1}$ is the Gaussian part of $\zeta$.
We see that the second order coefficient which is, apart from the numerical factor, nothing more than $f_{\rm NL}$ contains an enhancement
factor $1/s$~\footnote{Strictly speaking, at second order perturbation,
$\delta_1^2$ terms should also appear in (\ref{quasi-second}).
However, such terms only yield $f_{\rm NL}$ of ${\cal O}(1)$ and can be safely neglected when $s \ll 1$.}.
This mechanism to get large $f_{\rm NL}$ is exactly the same as the single curvaton case 
in which $f_{\rm NL}$ is inversely proportional to a fraction of the curvaton energy density at the
time when it decays into the radiation \cite{Lyth:2002my}.
 
Since $f_{\rm NL}$ is bounded to be $|f_{\rm NL}| \lesssim 100$ from the observations \cite{Komatsu:2010fb},
$s$ cannot be smaller than $10^{-2}$ if $\delta_2=0$.
Correspondingly, the observationally allowed maximum curvature perturbation 
when it is enhanced is at most $10^{-5}/s \simeq 10^{-3}$.
However, things change when $\delta_2$ is also allowed to take non-zero amplitude.
From (\ref{twofnl}), we find that if fluctuations in $\sigma_2$
contribute more to the final curvature perturbation 
than the $\sigma_1$-field fluctuations, {\it i.e.} $r \gtrsim 1$, then $f_{\rm NL}$ is suppressed by a factor $r^{-4}$
compared to a case with $r=0$.
Therefore, $s$ smaller than $10^{-2}$ can satisfy the bound $|f_{\rm NL}| \lesssim 100$ if $r$ is suitably chosen.
In particular, if $r \gtrsim s^{-1/4} \gg 1$, then $f_{\rm NL}$ becomes as small as ${\cal O}(1)$.
At this level of $f_{\rm NL}$,  non-linear evolutionary
effects become important and it will not be easy
to extract primordial $f_{\rm NL}$ from observations.
Interestingly, in this case, we find from (\ref{twotnl}) and (\ref{twognl}) that 
$\tau_{\rm NL},~g_{\rm NL} \gtrsim {\cal O}(s^{-1/2}) \gg 1$.
Therefore, strong non-Gaussianity appears in the trispectrum but not in the bispectrum \footnote{It is generally true that
the trispectrum becomes relativily stronger than the bispectrum if more than one field
contribute to the curvature perturbations. 
The local-type single field model yields a relation $\tau_{\rm NL} = \frac{36}{25} f_{\rm NL}^2$.
In \cite{Suyama:2007bg}, it was shown that an inequality 
$\tau_{\rm NL} \ge \frac{36}{25} f_{\rm NL}^2$ holds for any local-type multi-field model. In \cite{Ichikawa:2008iq},
a scenario that both inflaton and curvaton contribute to the curvature perturbations was considered.
It was shown that $\tau_{\rm NL}$ is enhanced by a factor $1+r^2$ compared to $\frac{36}{25} f_{\rm NL}^2$, where
$r$ represents the contribution of the inflaton fluctuations to the curvature perturbations.}.
In such a case, the trispectrum would be useful to search for non-Gaussianity.

From (\ref{twotnl}) and (\ref{twognl}), we can also derive a unique relation between $\tau_{\rm NL}$ and $g_{\rm NL}$ as
\begin{equation}
\frac{\tau_{\rm NL}}{g_{\rm NL}}=\frac{27}{5}+{\cal O} \left( \frac{1}{f_{\rm NL}} \right), \label{consist}
\end{equation}
which will be useful to observationally discriminate the
two curvaton model from the other models that generate
large non-Gaussianity.
Consistency relations between $f_{\rm NL}$ and $g_{\rm NL}$, $f_{\rm NL}$ and $\tau_{\rm NL}$
or among the three parameters have been obtained for various models.
As far as we know, (\ref{consist}) is the first example that gives the unique relation between $\tau_{\rm NL}$
and $g_{\rm NL}$ without using $f_{\rm NL}$ (for consistency relations for other models,
see \cite{Suyama:2010uj}).

\subsection{Stochastic gravitational waves}
It is well known that second order tensor perturbations are induced
by the first order scalar perturbations by the mode-mode couplings.
This means, $\Omega_{\rm GW}$, the energy density of GWs per unit
logarithmic interval of frequency, is proportional to quartic of the scalar perturbations,
motivating us to consider the generation of GWs in the two-curvaton model.

A basic picture of the production of GWs in the two curvaton model is that the scalar perturbations are 
temporarily enhanced equally on all the superhorizon scales and large magnitude of $\Omega_{\rm GW}$ with 
a frequency equal to the Hubble parameter is produced at each time while the enhanced modes are re-entering the horizon.
Therefore, we expect that the resulting $\Omega_{\rm GW}$ will have a broad peak of an interval of frequencies 
whose corresponding modes re-enter the horizon during the curvature perturbation is being enhanced. 
To get this kind of result, we have to solve the equation of motion for the second order
tensor perturbations as well as the perturbation equations for the scalar modes and
the background equations.
Although all the numerical calculations to arrive at the final result are straightforward,
it takes a long computation time due to multi-integrations.
To avoid this, we stick ourselves to cases where $\sigma_1$-field immediately
decays soon after it dominates the total energy density.
In other words, we assume that the epoch B terminates 
in a moment and the effects of the enhancement of 
the curvature perturbation shows up only at the epoch C which is radiation dominated.
In such cases, instead of numerically solving the perturbation equations for the scalar modes,
we can assume that the scalar modes which are to re-enter the horizon during the epoch C are 
already enhanced since the epoch A and can use the analytic transfer function for the 
radiation dominated universe to evolve the scalar modes.
By these assumptions, we do not need to numerically integrate the background equations and the
linearized equations for the scalar modes,
which drastically diminishes the task of the numerical computations. 
What is then left is to solve the evolution equations for the second order tensor perturbations 
sourced by the first order scalar perturbations.

In what follows, we first briefly review the general formalism to calculate $\Omega_{\rm GW}$ induced 
by the scalar-scalar couplings \cite{Ananda:2006af,Baumann:2007zm,Saito:2008jc} and then provide our results in two curvaton case.
We basically follow the notations of \cite{Saito:2009jt}.
The metric that is relevant to our purpose is given by the metric (\ref{scalarper}) plus the second
order tensor perturbations:
\begin{equation}
ds^2=a^2(\eta) \big[ -(1+2 \psi) d\eta^2+\{ (1-2 \phi)\delta_{ij} +h_{ij} \} dx^i dx^j \big],
\end{equation}
where $h_{ij}$ are the second order tensor perturbations which satisfy the transverse-traceless conditions:
\begin{equation}
h_{ij,j}=\delta^{ij} h_{ij}=0.
\end{equation}
In the following, we set $\psi=\phi$.
Since the gravitational waves are transverse waves, we can Fourier-decompose $h_{ij}$ as
\begin{equation}
h_{ij}(\eta,{\vec x})=\int \frac{d^3 k}{{(2\pi)}^3}~e^{i {\vec k} \cdot {\vec x}} \left( e_{ij} ({\vec k}) h_{\vec k}(\eta)+{\bar e}_{ij} ({\vec k}) {\bar h}_{\vec k} (\eta) \right),
\end{equation}
where $e_{ij} ({\vec k})$ and ${\bar e}_{ij}({\vec k})$ are the polarization tensors
orthogonal to ${\vec k}$. Introducing the unit vectors $e_i ({\vec k})$ and ${\bar e}_i ({\vec k})$
orthogonal to ${\vec k}$,
they are given by
\begin{equation}
e_{ij} ({\vec k})=\frac{1}{\sqrt{2}} \left( e_i ({\vec k})e_j ({\vec k})-{\bar e}_i ({\vec k}) {\bar e}_j ({\vec k}) \right), ~~~~~{\bar e}_{ij} ({\vec k})=\frac{1}{\sqrt{2}} \left( e_i ({\vec k}) {\bar e}_j ({\vec k})+{\bar e}_i ({\vec k}) e_j ({\vec k}) \right).
\end{equation}
Then the evolution equation for $h_{\vec k}$ is given by
\begin{equation}
h_{\vec k}^{''}+2 {\cal H} h_{\vec k}^{'}+k^2 h_{\vec k}=S_{\vec k}, \label{EOM:h}
\end{equation}
where the source term is given by
\begin{equation}
S_{\vec k}=2 e^{ij} ({\vec k}) \int \frac{d^3 q}{{(2\pi)}^3} ~q_\ell q_m \left( 3 \phi_{\vec q} \phi_{{\vec k}-{\vec q}}+\frac{2}{\cal H} \phi_{\vec q} \phi_{{\vec k}-{\vec q}}^{'}+\frac{1}{{\cal H}^2} \phi_{\vec q}^{'} \phi_{{\vec k}-{\vec q}}^{'} \right). 
\end{equation}
In the radiation dominated universe, the solution of Eq.~(\ref{EOM:h}) is given by
\begin{equation}
u_{\vec k}(\eta)=\int_{-\infty}^\eta d\eta'~g_k (\eta,\eta') a(\eta') S_k (\eta'),
\end{equation}
where $u_{\vec k} = a h_{\vec k}$ and $g_k (\eta,\eta')$ is the retarded Green's function:
\begin{equation}
g_k (\eta,\eta')=\frac{1}{k} \sin  k(\eta-\eta').
\end{equation}
Then the two-point function of $h_{\vec k}$ becomes
\begin{equation}
\langle h_{\vec k_1} h_{\vec k_2} \rangle \\
=\frac{1}{a^2(\eta)}\int_{-\infty}^\eta d\eta_1~\int_{-\infty}^\eta d\eta_2~g_{k_1} (\eta,\eta_1) g_{k_2} (\eta,\eta_2)a(\eta_1)a(\eta_2) \langle S_{\vec k_1}(\eta_1) S_{\vec k_2}(\eta_2) \rangle.
\end{equation}
We need to evaluate $\langle S_{\vec k_1} (\eta_1) S_{\vec k_2}(\eta_2) \rangle$.
To this end, we need time evolution of $\phi_{\vec k}$. 
Its evolution in the radiation dominated universe is given by
\begin{equation}
\phi_{\vec k}(\eta)=D_k(\eta) \phi_0({\vec k}), ~~~~~D_k(\eta)=\frac{9}{{(k\eta)}^2} \left( \frac{\sqrt{3}}{k\eta} \sin \left( \frac{k\eta}{\sqrt{3}} \right)-\cos  \left( \frac{k\eta}{\sqrt{3}} \right) \right).
\end{equation}
Using this, the two-point function of the source can be written as
\begin{eqnarray}
\langle S_{\vec k_1} (\eta_1) S_{\vec k_2}(\eta_2) \rangle= &&8 e_{ij} ({\vec k_1}) e_{mn} (-{\vec k_1}) {(2\pi)}^3 \delta ({\vec k_1}+{\vec k_2}) \int \frac{d^3q}{{(2\pi)}^3} P_\phi (|{\vec k_1}-{\vec q}|) P_\phi (q) \nonumber \\
&& \times q_i q_j q_m q_n f(|{\vec k_1}-{\vec q}|,q,\eta_1)f(|{\vec k_1}-{\vec q}|,q,\eta_2),
\end{eqnarray}
where 
\begin{eqnarray}
&&\langle \phi_0 ({\vec k_1}) \phi_0 ({\vec k_2}) \rangle = {(2\pi)}^3 P_\phi (k_1) \delta ({\vec k_1}+{\vec k_2}), \\
&&f(k_1,k_2,\eta) \equiv 3 D_{k_1}(\eta) D_{k_2}(\eta)+\eta \left( 2 D_{k_1}(\eta)+\eta D_{k_1}^{'} (\eta) \right) D_{k_2}^{'} (\eta). \label{eq:f}
\end{eqnarray}

We define the power spectrum of the tensor perturbation by 
\begin{equation}
\langle h_{\vec k_1}(\eta) h_{\vec k_2}(\eta) \rangle ={(2\pi)}^3 P_h (\eta,k_1) \delta ({\vec k_1}+{\vec k_2})
\end{equation}
When the source term can be neglected, the energy density of the tensor perturbation can be
written in terms of their power spectrum well inside the horizon,
\begin{equation}
\Omega_{\rm GW}(\eta,k)=\frac{1}{6} {\left( \frac{k}{\cal H} \right)}^2 {\cal P}_h (\eta,k),
\end{equation}
where ${\cal P}_h \equiv k^3 P_h/{(2\pi)}^2$ is the dimensionless power spectrum.

Given the general formalism to calculate $\Omega_{\rm GW}$, 
we are ready to apply it to the two curvaton model.
Due to the reasoning we mentioned at the beginning of this subsection, 
we use top-hat type function with a width $\Delta$ for ${\cal P}_\phi (k)$:
\begin{equation}
{\cal P}_\phi(k)=
	\begin{cases}
		{\displaystyle \frac{{\mathcal A}^2}{2\Delta}} & \text{for}~~|\ln (k/{k_p})|<\Delta , \\
		0 & \text{otherwise},
	\end{cases}
\end{equation}
where ${\cal P}_\phi \equiv k^3 P_\phi/{(2\pi)}^2$, $k_p$ is the peak wavenumber and ${\mathcal A}^2$ corresponds to the total power of the spectrum.
Then $\Omega_{\rm GW}$ can be written as \cite{Saito:2009jt}
\begin{equation}
\Omega_{\rm GW}(\eta,k)=\frac{k}{6} \int_{q_{\rm min}}^{e^\Delta} dq \int_{-\delta_{\rm max}}^{\delta_{\rm max}} d\delta~{\cal F}^2(k,\eta,q,\delta),
\end{equation}
where 
\begin{equation}
q_{\rm min}={\rm max} \{ k/2,e^{-\Delta} \},~~~\delta_{\rm max}={\rm min} \bigg\{ 1,\frac{2}{k}\sinh \Delta \bigg\}.
\end{equation}
The function ${\cal F}(k,\eta,q,\delta)$ is defined by
\begin{equation}
{\cal F}(k,\eta,q,\delta)=\frac{k}{2\Delta} \frac{\left( 4q^2-k^2 \right) (1-\delta^2)}{4q^2-k^2 \delta^2} I\left( k,q+\frac{k\delta}{2},q-\frac{k\delta}{2},\eta \right), 
\end{equation}
where
\begin{equation}
I(k,k_1,k_2,\eta) =k \int_0^\eta d\eta_1~a(\eta_1) g_k(\eta,\eta_1) f(k_1,k_2,\eta_1).
\end{equation}

\begin{figure}[t]
  \begin{center}{
    \includegraphics{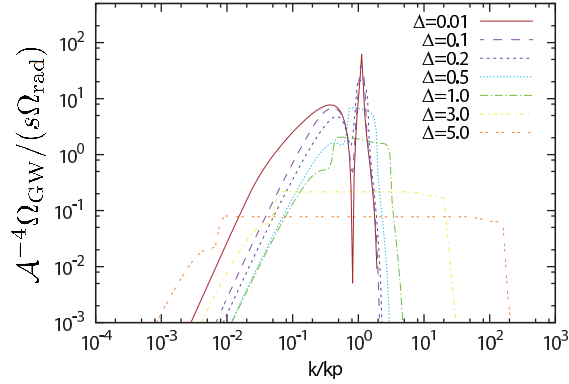}
    }
  \end{center}
  \caption{$\Omega_{\rm GW}(k,\eta)$ for various $\Delta$. The time is chosen to be $\eta=10^3/k$.}
 \label{fig:OmegaGW}
\end{figure}

In Fig.~\ref{fig:OmegaGW}, we show numerically calculated ${\cal A}^{-4} \Omega_{\rm GW}(k,\eta)/(s \Omega_{\rm rad})$ for various values
of $\Delta$ \footnote{The results presented here look somewhat
different from the ones given in \cite{Saito:2009jt}
which also obtained $\Omega_{\rm GW}$ for the top-hat type power
spectrum, although the qualitative
features are the same. This is due to there were a couple of mistakes in 
the manipulation in \cite{Saito:2009jt}.}.
A suppression factor $s$ is due to the decay of $\Omega_{\rm GW}$ during the epoch D where the
universe expands like matter dominated regime and the other one $\Omega_{\rm rad} \simeq 8\times 10^{-5}$ comes from
the matter dominated universe after the matter-radiation equality.
We see that for $\Delta \lesssim 0.5$, $\Omega_{\rm GW}$ has a strong peak at $k=k_p$ 
and decays as $k^{-3}$ for $k \ll k_p$.
In all the cases, $\Omega_{\rm GW}$ has a sharp drop at $k/k_p \simeq e^\Delta$.
This is because the momentum conservation prohibits a generation of the tensor mode whose wavenumber is
greater than $2k_p e^\Delta$.
For $\Delta \gtrsim 0.5$, $\Omega_{\rm GW}$ has plateau between a range $e^{-\Delta} \lesssim k/k_p \lesssim e^\Delta$
and decays as $k^{-3}$ for $k \lesssim e^{-\Delta} k_p$.
The magnitude of the plateau is well fitted with $\simeq 2.0 \times \Delta^{-2}$.
Therefore, the magnitude of the plateau at present is given by
\begin{equation}
\Omega_{\rm GW,plateau} \simeq \frac{2 {\cal A}^4}{\Delta^2}s \Omega_{\rm rad} \simeq 2\times 10^{-17} {\left( \frac{\cal A}{10^{-2}} \right)}^4 \left( \frac{s}{10^{-4}} \right) {\left( \frac{\Delta}{3} \right)}^{-2}. \label{omega-plateau}
\end{equation}
The current frequency corresponding to the upper limit of the plateau is
\begin{equation}
f_{\rm max} \simeq \frac{\sqrt{\Gamma_1 H_{\rm eq}}}{2\pi z_{\rm eq}}s^{1/4} \simeq 10~{\rm Hz}~ {\left( \frac{s}{10^{-4}} \right)}^{1/4} {\left( \frac{\Gamma_1}{100~{\rm GeV}} \right)}^{1/2}, \label{frequency-max}
\end{equation}
where $z_{\rm eq} \simeq 3200$ and $H_{\rm eq}$ are the redshift and the Hubble
parameter at the time of the matter-radiation equality respectively.
The lowest frequency is 
\begin{equation}
f_{\rm min} = {\left( \frac{\Omega_{2,*}}{\Omega_{1,*}} \right)}^2 f_{\rm max}.
\end{equation}

\begin{figure}[t]
  \begin{center}{
    \includegraphics[width=15cm]{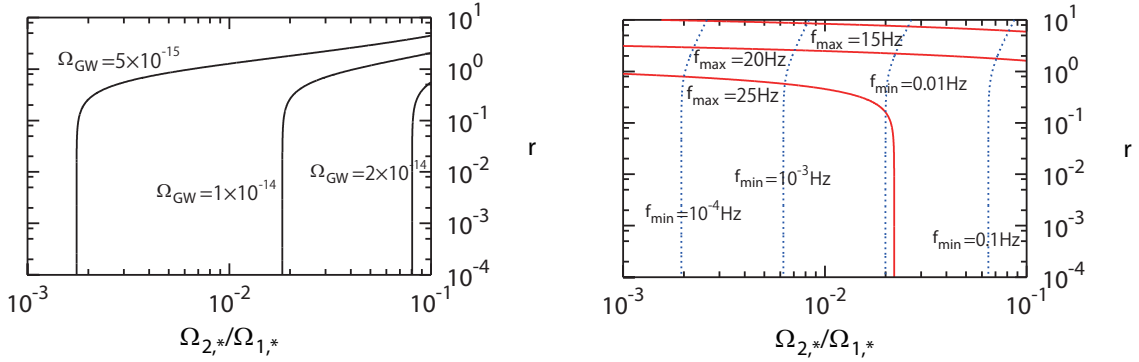}
    }
  \end{center}
  \caption{Left panel shows contours of $\Omega_{\rm GW,plateau}$ and the right one shows
  contours of $f_{\rm max}$ and $f_{\rm min}$. The other free parameters ${\cal A}$ and $\Gamma_1$ 
  are fixed to be ${\cal A}^2=5\times 10^{-4}$ and $\Gamma_1=100 ~{\rm GeV}$.}
 \label{fig:combined}
\end{figure}

\begin{figure}[t]
  \begin{center}{
    \includegraphics[width=8cm]{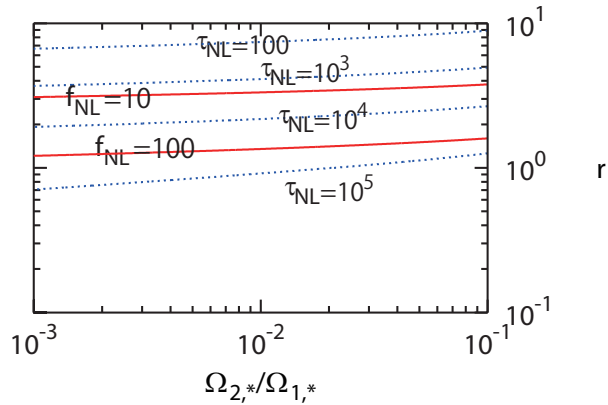}
    }
  \end{center}
  \caption{This panel shows contours of $f_{\rm NL}$ and $\tau_{\rm NL}$. The other free parameters ${\cal A}$ and $\Gamma_1$ 
  are fixed to be ${\cal A}^2=5\times 10^{-4}$ and $\Gamma_1=100 ~{\rm GeV}$.}
 \label{fig:NG}
\end{figure}

\begin{figure}[t]
  \begin{center}{
    \includegraphics[width=12cm]{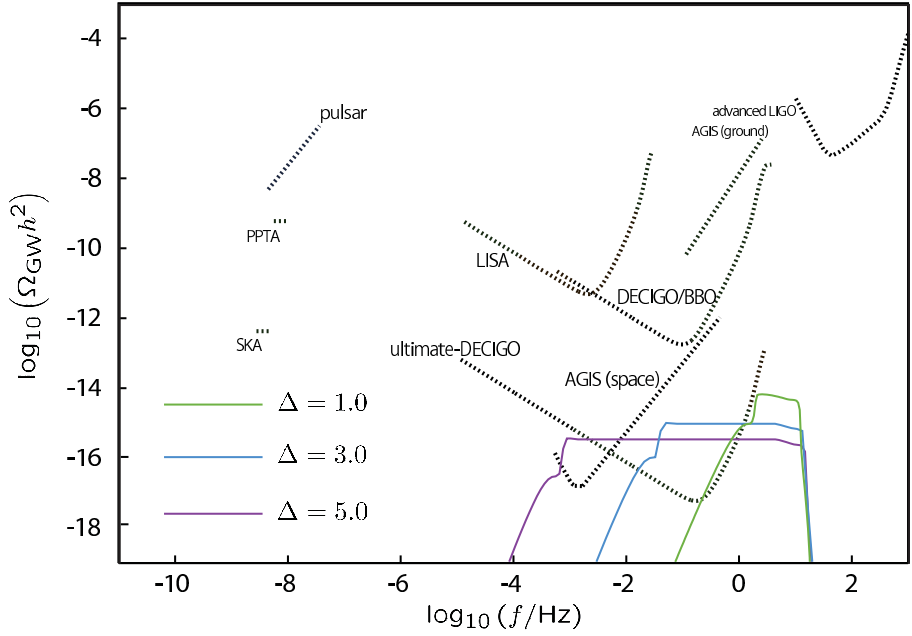}
    }
  \end{center}
  \caption{This panel show plots of $\Omega_{\rm GW}$ for three cases:
$\Delta=1.0,~3.0,~5.0$, assuming $\Gamma_1=100 ~{\rm GeV},~{\cal A}^2 = 5\times 10^{-4}$ and
$r=3$ with expected sensitivity of the future GW detectors such as LISA, DECIGO/BBO and AGIS.
}
 \label{limit}
\end{figure}

It is interesting to derive an upper bound on $\Omega_{\rm GW,plateau}$.
From the expression of $f_{\rm NL}$ given by Eq.~(\ref{twofnl}) and the WMAP normalization \cite{Komatsu:2010fb},
\begin{equation}
{\cal P}_\phi (\eta_f) = (1+r^2) s^2 \frac{{\cal A}^2}{2 \Delta} \simeq 10^{-9}, \label{wmap-normalization}
\end{equation}
we can write $s$ and $r$ in terms of $f_{\rm NL}$ and ${\cal A}^2/\Delta$.
By using these relations, we find that $\Omega_{\rm GW,plateau}$ is bounded from above as
\begin{equation}
\Omega_{\rm GW,plateau} \simeq 5 \times 10^{-10} {\left( \frac{{\cal A}^2}{2\Delta} \right)}^{4/3} f_{\rm NL}^{1/3} < 10^{-13},
\end{equation}
where in the last inequality, we used $\Delta \gtrsim 0.5,~f_{\rm NL} \lesssim 50$ and ${\cal A}\lesssim 0.05$.
As we will see in the next subsection, larger value than ${\cal A}=0.05$ over-produces 
black holes, which is excluded by the observations.
Therefore, $\Omega_{\rm GW,plateau}=10^{-13}$ is the possible maximal amplitude that can be achieved
in principle.
This amplitude is much smaller than the one given in \cite{Saito:2009jt}.
This is mainly because $\Omega_{\rm GW,plateau}$ in the two curvaton model considered here is multiplied by
the suppression factor $s$ (see Eq.~(\ref{omega-plateau})) due to the existence of matter dominance epoch (epoch D).
On the other hand, \cite{Saito:2009jt} does not consider such an epoch and hence there is
no additional suppression on $\Omega_{\rm GW}$. 
As can be seen from Fig.~\ref{limit}, $\Omega_{\rm GW,plateau}=10^{-13}$ is slightly lower than the sensitivity of 
LISA \cite{LISA} and DECIGO/BBO \cite{Seto:2001qf,BBO}.
But it is much higher than the sensitivities achieved by ultimate-DECIGO and space-based AGIS \cite{Dimopoulos:2008sv}.

As an example, we show contour plots of $f_{\rm max},~f_{\rm min}$
and $\Omega_{\rm GW,plateau}$ in Fig.~\ref{fig:combined} and $f_{\rm NL}$ and $\tau_{\rm NL}$
in Fig.~\ref{fig:NG} as functions of $r$ defined by Eq.~(\ref{def-r}) and $\Omega_{2,*}/\Omega_{1,*}$.
The other free parameters ${\cal A}$ and $\Gamma_1$ are fixed to be ${\cal A}^2=5\times 10^{-4}$ and
$\Gamma_1=100 ~{\rm GeV}$. $\Delta$ is determined by a relation
\begin{equation}
\Delta = \log \left( \frac{\Omega_{1,*}}{\Omega_{2,*}}\right),
\end{equation}
and $s$ is determined by the WMAP normalization (\ref{wmap-normalization}).

We find that $\Omega_{\rm GW,plateau}$ is ${\cal O}(10^{-14})$ in the frequency band $10^{-4}~{\rm Hz} - 30~{\rm Hz}$. 
We also find that in this case, $r \lesssim 1$ is ruled out by the observational bound on $f_{\rm NL}$,
$|f_{\rm NL}| < 100$.
This is because $s$ is typically $10^{-3}$ in this case and we have $f_{\rm NL} ={\cal O}(10^3)$ for
$r \lesssim 1$.
As we have seen in the previous subsection, $\tau_{\rm NL}$ becomes very large as ${\cal O}(10^3)-{\cal O}(10^4)$.
Hence the strong non-Gaussian signal appears in the trispectrum rather than in the bispectrum.

In Fig.~\ref{limit}, we show plots of $\Omega_{\rm GW}$ for three cases:
$\Delta=1.0,~3.0,~5.0$, assuming $\Gamma_1=100 ~{\rm GeV},~{\cal A}^2 = 5\times 10^{-4}$ and
$r=3$ with expected sensitivity of the future GW detectors such as LISA, DECIGO/BBO and AGIS.
As mentioned before, we see that maximum of $\Omega_{\rm GW}$ is below the LISA and DECIGO/BBO sensitivities.
On the other hand, ultimate-DECIGO and space-based AGIS will be able to probe GWs generated 
in two curvaton models if the model parameters are suitably chosen.
If we lower $\Gamma_1$ as small as $0.1~{\rm meV}$, then $\Omega_{\rm GW,plateau}$ does not
change but $f_{\rm max}$ becomes ${\cal O}(10^{-6}~{\rm Hz})$.
Hence $\Omega_{\rm GW,plateau}$ enters the frequency region of the pulsar timing.
However, as is clear from Fig.~\ref{limit}, $\Omega_{\rm GW,plateau}$ is far below the upper limit
coming from the pulsar timing observations.

\subsection{Primordial black holes}
PBHs are formed if there exist density fluctuations of the order unity and when such modes
re-enter the Hubble radius \cite{Hawking:1971ei,Carr:1974nx}.
The mass of the PBHs is roughly equal to the horizon mass at the time of formation,
\begin{equation}
G M_{\rm PBH} \simeq \frac{1}{H}. \label{pbh-mass}
\end{equation}
To simplify the analysis and also to make it consistent with the study of the last subsection,
we again assume that the epoch B terminates in a moment.
Then, PBHs would be efficiently formed during the epoch C and the mass range of the resultant PBHs is estimated as
\begin{equation}
\frac{1}{\Gamma_1} \lesssim G M_{\rm PBH} \lesssim \frac{1}{\Gamma_1} {\left( \frac{\Omega_{1,*}}{\Omega_{2,*}}\right)}^2. \label{mass-range}
\end{equation}

Let us write the fraction of the energy density of PBHs of mass between $(M,M+dM)$ 
at the time of formation as
\begin{equation}
\frac{d\beta_{\rm PBH}}{dM} dM.
\end{equation}
Then the total fraction of the energy density of PBHs is given by the integral of the above
quantity,
\begin{equation}
\beta_{\rm PBH} = \int dM~\frac{d\beta_{\rm PBH}}{dM}, \label{omega-pbh}
\end{equation}
where the range of integration is given by Eq.~(\ref{mass-range}).
The purpose of this subsection is to provide $\beta_{\rm PBH}$ for the two curvaton model
and to discuss its cosmological implications.
Since most of the necessary formulae and their detailed derivations are written in \cite{Saito:2009jt}, 
we omit the intermediate calculations and provide only results.

Assuming that the gravitational potential $\phi$ smoothed over the horizon size is Gaussian,
$d\beta_{\rm PBH}/dM$ can be written as
\begin{equation}
\frac{d \beta_{\rm PBH}}{dM}dM= {\cal P}_\phi (R_M^{-1}) \frac{\phi_c}{2 \sqrt{2 \pi} \sigma_{R_M}^3} \exp \left( -\frac{\phi_c^2}{2 \sigma_{R_M}^2} \right) \frac{dM}{M},
\end{equation}
where $R_M$ is the comoving horizon length, {\it i.e.} $a R_M = GM$, with $a$ being the scale factor
at the time of horizon crossing, $\phi_c$ is the threshold value of $\phi$ for black hole formation and 
\begin{equation}
\sigma_{R_M}^2 \equiv \int_0^{R_M^{-1}} \frac{dk}{k}~{\cal P}_\phi (k),
\end{equation}
is the variance of $\phi$ smoothed over the horizon size.
Although $\phi_c$ depends on the initial configuration of the perturbations \cite{Shibata:1999zs,Polnarev:2006aa},
we simply use $\phi_c = 0.5$ \cite{Carr:1975qj}.

For the top-hat type of ${\cal P}_\phi (k)$ which we have considered in this paper,
$\sigma_{R_M}$ becomes
\begin{equation}
\sigma_{R_M}^2= \frac{ {\cal A}^2}{2 \Delta} \log \left( \frac{e^\Delta}{k_p R_M} \right).
\end{equation}
Having these, we are ready to do the integration (\ref{omega-pbh}). The result is
\begin{equation}
\beta_{\rm PBH}\simeq \sqrt{\frac{2}{\pi}} \frac{\Delta \phi_c^2}{\cal A} \exp \left(-\frac{\phi_c^2}{2{\cal A}^2}\right). \label{omega-pbh2}
\end{equation}

\begin{figure}[t]
  \begin{center}{
    \includegraphics[width=8cm]{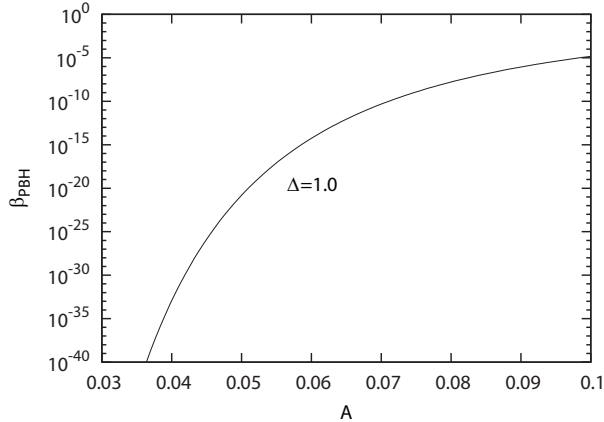}
    }
  \end{center}
  \caption{This panel shows $\beta_{\rm PBH}$ as a function of ${\cal A}$. $\Delta$
  is chosen to be $\Delta=1$.}
 \label{beta-PBH}
\end{figure}

Fig.~\ref{beta-PBH} shows $\Omega_{\rm PBH}h^2$ as a function of ${\cal A}$. 
$\Delta$ is chosen to be $\Delta=1.0$.
It is clear that $\beta_{\rm PBH}$ is very sensitive to change of ${\cal A}$.
Since various cosmological observations put different constraints on the abundance of PBHs 
for different mass,
using $\beta_{\rm PBH}$ to constrain ${\cal A}$ is not correct in a precise sense.
However, because of the strong sensitivity of $\beta_{\rm PBH}$ to ${\cal A}$,
even the change of many order of magnitude of $\beta_{\rm PBH}$ corresponds to 
the change of a factor of a few in ${\cal A}$.
This fact validates the use of $\beta_{\rm PBH}$ to constrain ${\cal A}$ as a first
approximation.
According to \cite{Carr:2009jm}, the constraint on $\beta_{\rm PBH}$ varies from
$10^{-30}$ to $10^{-10}$.
This can be converted to a constraint on ${\cal A}$ as ${\cal A} \lesssim 0.05$.

Since $\sigma_{R_M}$ is a monotonically decreasing function of $M$,
the mass of PBHs that dominantly contribute to (\ref{omega-pbh2}) is the lower limit
of Eq.~(\ref{mass-range}), which is estimated as
\begin{equation}
M_{\rm PBH} \simeq 8 \times 10^{12}~{\rm g} {\left( \frac{\Gamma_1}{100~{\rm GeV}} \right)}^{-1}.
\end{equation}
Combining this with Eq.~(\ref{frequency-max}) gives a relation between BH mass and
the corresponding frequency of GWs,
\begin{equation}
M_{\rm PBH} \simeq 1.5\times 10^{15}~{\rm g}~{\left( \frac{f_{\rm max}}{10~{\rm Hz}} \right)}^{-2} {\left( \frac{s}{10^{-4}} \right)}^{1/2}. 
\end{equation}

\section{Summary}
If more than one curvaton dominates the Universe at different epochs,
the curvature perturbations can be temporarily enhanced to a value much
larger than the observed one $10^{-5}$.
In this paper, we studied in detail the evolution of the curvature perturbations
in two curvaton models.
By solving the linearized perturbation equations both numerically and analytically, we confirmed
that the curvature perturbations are indeed enhanced during the period from
the time when the first decaying curvaton dominates the Universe until
the second decaying curvaton dominates the Universe.
The amplitude of the enhanced curvature perturbation is roughly equal to
the density perturbation of the first decaying curvaton.
We then provided an analytic expression of the full-order curvature perturbation 
which does not rely on the sudden decay approximation and is exact on super-horizon scales.
At the linear order, we compared with the analytic results with the numerical ones
and found they agree very well.

The temporal enhancement of the curvature perturbations leaves its traces as
the strong non-Gaussian perturbations, stochastic gravitational waves generated by 
the scalar-scalar mode couplings and the primordial black holes.
By using the analytic formula for the full-order curvature perturbations,
we gave the expressions for the so-called non-linearity parameters $f_{\rm NL}, \tau_{\rm NL}$
and $g_{\rm NL}$.
If both two curvatons contribute to the final curvature perturbations, then
the strongest non-Gaussian signal comes from $\tau_{\rm NL}$ and $g_{\rm NL}$
rather than from $f_{\rm NL}$. 
We also gave a consistency relation between $\tau_{\rm NL}$ and $g_{\rm NL}$ without
using $f_{\rm NL}$.
If non-Gaussianity is detected in the future,
this relation would be useful to discriminate this model from the others that also
generate large non-Gaussian perturbations.

We next studied the generation of GWs sourced by the enhanced curvature perturbations.
The spectrum of $\Omega_{\rm GW}$ has a plateau corresponding to the duration of 
the enhancement.
Because of the existence of the period where the secondly decaying curvaton dominates
the Universe and the Universe expands like the matter dominated universe,
$\Omega_{\rm GW}$ today is accompanied by a suppression factor which represents the 
fraction of the radiation energy density coming from the first decaying curvaton
at the time when the secondly decaying curvaton decays.
Due to this suppression factor, the possible maximal amplitude of $\Omega_{\rm GW}$ at 
the plateau is at most $10^{-13}$, which is below the LISA and DECIGO/BBO sensitivities,
but above the ultimate-DECIGO and space-based AGIS sensitivities.
Actually, if the decay rate of the first decaying curvaton is around $100~{\rm GeV}$,
then the frequency interval of the plateau can be $10^{-3}~{\rm Hz} \sim 10~{\rm Hz}$.
Such a case can be a target of the ultimate-DECIGO and AGIS.

We finally calculated the abundance of PBHs which are formed by the gravitational
collapse of the enhanced curvature perturbations.
We then provided an upper bound on the amplitude of the enhanced curvature perturbation
by using the observational upper bounds on the abundance of PBHs.\\

\noindent {\bf Acknowledgments:} 
TS would like to thank Christophe Ringeval, Ryo Saito, Tomo Takahashi
and Shuichiro Yokoyama for useful comments.
This work was partially supported by a Grant-in-Aid for JSPS Fellows
No.~1008477(TS), JSPS  Grant-in-Aid for Scientific Research
No.\ 23340058 (JY), and the Grant-in-Aid for
Scientific Research on Innovative Areas No.\ 21111006 (JY).

\appendix


\section{Derivation of Eq.\,(\ref{full-efolding})}
\label{app:1}
In this appendix, we derive Eq.\,(\ref{full-efolding}).
In the following, we frequently use epoch A, epoch B, epoch C and epoch D which are defined in Sec.\ref{sec.basic}.
For convenience, we use the cosmic time $t$ as a time variable.
The basic equations we use are the background equations (\ref{backa}),
 (\ref{backb})
for the curvatons and   (\ref{backr}) for the radiation.
In the spirit of the $\delta N$ formalism, the solutions of these
equations, (\ref{rhoone}), (\ref{rhotwo}), and (\ref{rhor})
 depend on the space coordinate ${\vec x}$
and evolve independently of other points,
although we do not show their dependence explicitly. 
Also, $t$ should be interpreted as the local proper time at ${\vec x}$ hereafter.

If we choose $t$ such that $\Gamma_2 t \gg 1$, {\it i.e.}, an epoch well after the $\sigma_2$-field decay,
then $\rho_{\rm rad}$ can be safely replaced by the total energy density
$\rho(t)$.
Then from (\ref{rhor}) we find
\begin{equation}
a^4 (t) \rho(t)=\int_0^t dt' \, \left( \Gamma_1 a(t') a_*^3 \rho_{1,*} e^{-\Gamma_1 t'}+\Gamma_2 a(t') a_*^3 \rho_{2,*} e^{-\Gamma_2 t'} \right).
\end{equation}
Also since the integrand is exponentially suppressed for $t \gg 1/\Gamma_2$, we can push the upper limit of integration to infinity.
Then, the e-folding number from the initial time to the time when the total energy density becomes $\rho_f$ is given by
\begin{equation}
N=\frac{1}{4} \log \left( \Gamma_1 \Omega_{1,*}F_1+\Gamma_2 \Omega_{2,*}F_2 \right)+\frac{1}{4} \log \left( \frac{\rho_*}{\rho_f} \right), \label{app1}
\end{equation}
where $F_1$ and $F_2$ are defined by
\begin{align}
&F_1 \equiv \int_0^\infty dt\,a(t) e^{-\Gamma_1 t}, \label{defFa}\\
&F_2 \equiv \int_0^\infty dt\,a(t) e^{-\Gamma_2 t}. \label{defFb}
\end{align}
We see that the e-folding number depends on the field values of $a$ and $b$ through $F_1$ and $F_2$
as well as $\Omega_{1,*}$ and $\Omega_{2,*}$.
Now the problem is reduced to deriving the analytic expression of $F_1$ and $F_2$.

Let us first evaluate $F_1$.
To evaluate it, we need to know the evolution of the scale factor.
Since the integrand of $F_1$ has an exponential cutoff for $t \gg 1/\Gamma_1$, 
the dominant contribution to the integral comes from a time interval where $\sigma_1$-field dominates the universe (epoch B).
During that epoch, we can neglect the $\sigma_2$-field because its energy density is tiny compared to the total one.
Then, the background equations in terms of $\Omega_1$ and $H$ are given by
\begin{align}
&\frac{d \Omega_1}{dt}=\big\{ (1-\Omega_1) H-\Gamma_1 \big\} \Omega_1, \\
&\frac{dH}{dt}=-\frac{1}{2}(4-\Omega_1) H^2.
\end{align}
We want to solve these equations from the time $t=t_b$ when the $\sigma_1$-field is dominating the universe, but still well before
the $\sigma_1$-field decay, to the time well after the $\sigma_1$-field decay.
To know the magnitude of the scale factor at $t_b$,
we need to connect the scale factor in that epoch with the initial scale factor $a_*$.
We can make the connection by solving the differential equation for $a(t)$:
\begin{equation}
H^2=H_*^2 \bigg\{ {\left( \frac{a}{a_*} \right)}^{-4} (1-\Omega_{1,*})+{\left( \frac{a}{a_*} \right)}^{-3} \Omega_{1,*} \bigg\}.
\end{equation}
We can exactly integrate this equation. The result is
\begin{equation}
\frac{2}{3 \Omega_{1,*}^2} \bigg\{ 2-3 \Omega_{1,*}+ \sqrt{1+\left( \frac{a}{a_*}-1 \right)\Omega_{1,*} } \left( -2+\left( \frac{a}{a_*}+2\right) \Omega_{1,*} \right) \bigg\}=H_* (t-t_*).
\end{equation}
Neglecting $t_*$, we find that $a(t)$ at $t_b$ is given by
\begin{equation}
a(t_b) = a_* {\left( \frac{9}{4} \Omega_{1,*} \right)}^{1/3} {(H_* t_b)}^{2/3}+\frac{a_*}{\Omega_{1,*}}+{\cal O} \left( \frac{1}{t_b} \right). \label{appscalef}
\end{equation}
For large $t_b$, the first term dominates the scale factor. 
The second term represents the contribution from the radiation being diluted faster than the $\sigma_1$-field.
As we will see, keeping only the first term gives the desired expression of $F_1$.

Things become clearer if we introduce the new dimensionless variables by
\begin{align}
&b(t) \equiv \frac{a(t)}{a_* {\left( \frac{9}{4} \Omega_{1,*} \right)}^{1/3} {(H_* t)}^{2/3}}, \label{appdefb} \\
&E(t) \equiv \frac{3t}{2} H(t). \label{appdefE}
\end{align}
Defining the dimensionless time $s$ by $s\equiv \Gamma_1 t$, the evolution equations for the new variables are given by
\begin{align}
&\frac{db}{ds}=\frac{2b}{3s} (E-1), \label{appdb} \\
&\frac{dE}{ds}=\frac{E}{s} \big\{ 1-(4-\Omega_1) \frac{E}{3} \big\}, \label{appdE} \\
&\frac{d\Omega_1}{ds}=\big\{ \frac{2E}{3s}(1-\Omega_1)-1 \big\} \Omega_1. \label{appdomega}
\end{align}
The initial conditions are $b(0)=E(0)=\Omega_1(0)=1$.
The new set of differential equations plus the initial conditions are free of model parameters,
which means that we have extracted the dependence of $F_1$ on the model parameters.
Indeed, in terms of the new variables, $F_1$ can be written as
\begin{equation}
F_1=a_* {\left( \frac{9}{4} \Omega_{1,*} H_*^2 \right)}^{1/3} \frac{c_\Gamma}{\Gamma_1^{5/3}}, \label{appFa}
\end{equation}
where $c_\Gamma$ is a purely numerical value defined by
\begin{equation}
c_\Gamma \equiv \int_0^\infty ds \, s^{2/3} b(s) e^{-s}.
\end{equation}
We find numerically that $c_\Gamma \approx 0.830$.

Let us next evaluate $F_2$.
For $F_2$, the dominant contribution to the integral comes from a time interval when the $\sigma_2$-field dominates the universe (epoch D).
Since $\sigma_1$-field has completely decayed into the radiation by that epoch, we can only consider the radiation
and the $\sigma_2$-field in the following analysis.
To make an argument similar to the case of $F_1$,
first we have to connect the scale factor at the epoch C with $a_*$.
The scale factor at the epoch C is given by $b(s)$ for $s \gg 1$. 
From (\ref{appdb}) and (\ref{appdE}), we find that the asymptotic form of $b(s)$ for $s \gg 1$ is given by
\begin{equation}
b(s)=d_\Gamma s^{-1/6} \left( 1+{\cal O}(s^{-1}) \right),
\end{equation}
where $d_\Gamma$ is a numerical constant. We find numerically $d_\Gamma \approx 1.101$.
Substituting this into (\ref{appdefb}), the scale factor at a time $t=t_c$ in the epoch C is given by
\begin{equation}
a(t_c)=a_* {\left( \frac{9}{4} \Omega_{1,*} \right)}^{1/3} d_\Gamma \frac{H_*^{2/3}}{\Gamma_1^{1/6}} t_c^{1/2}. \label{appscaleC}
\end{equation}
Using this result, the $\Omega_2$ at this time is found to be
\begin{equation}
\Omega_2 (t_c)=\frac{16}{9} \frac{\Omega_{2,*}}{\Omega_{1,*}} \frac{\Gamma_1^{1/2}}{d_\Gamma^3} t_c^{1/2}. \label{appomegabc}
\end{equation}

Now it may be expected that the things can be exactly mapped to what we did for $F_1$ by replacing the
epoch A/B by the epoch C/D.
To be more precise, simply replacing $a_*,\,\Omega_{1,*},\,H_*$ and $\Gamma_1$ appearing in (\ref{appFa}) by $a(t_c)$ given by (\ref{appscaleC}), 
$\Omega_2$ given by (\ref{appomegabc}), $H(t_c)$ and $\Gamma_2$ would give the desired expression for $F_2$.
However, as it will come out, we need to include a term in the scale factor corresponding to the second term of (\ref{appscalef}) to have the correct answer,
which makes the calculations more complicated.

If we denote by $t_d$ the time when the $\sigma_2$-field is dominating the universe, but still well before the $\sigma_2$-field decay,
the scale factor at that time is given by 
\begin{equation}
a(t_d) = a(t_c) {\left( \frac{9}{4} \Omega_2 (t_c) H^2(t_c) \right)}^{1/3} t_d^{2/3}+\frac{a(t_c)}{\Omega_2(t_c)}+{\cal O} \left( \frac{1}{t_d} \right), \label{appscalef2}
\end{equation} 
which is obtained by exactly the same argument as what we used to derive (\ref{appscalef}).
After the example of $F_1$, let us introduce the new dimensionless variables by
\begin{align}
&d(t) \equiv \frac{a(t)}{a(t_c) {\left( \frac{9}{4} \Omega_2 (t_c) H^2(t_c) \right)}^{1/3} t^{2/3}}. \label{appdefd} 
\end{align}
To include the second term of (\ref{appscalef2}), let us decompose the scale factor as
\begin{equation}
a(t)={\bar a}(t)+\delta a(t).
\end{equation}
Here ${\bar a}(t) ( \delta a(t))$ is a part of the scale factor which reduces to the first(second) term in (\ref{appscalef2}) at $t_d$.
Correspondingly, we can define ${\bar d}(t),\,\delta d(t),\,{\bar E}(t),\,\delta E(t),\,{\bar \Omega_2}(t)$ and $\delta \Omega_2 (t)$.
The evolution equations for ${\bar d}(t),\,{\bar E}(t)$ and ${\bar \Omega_2}(t)$ are given
by (\ref{appdb})-(\ref{appdomega}) with $b,\,E,\,\Omega_1$ and $s$ being replaced by ${\bar d},\,{\bar E},\,{\bar \Omega_1}$
and $u \equiv \Gamma_2 t$.
The initial conditions for those variables are ${\bar d}(0)={\bar E}(0)={\bar \Omega_2}(0)=1$.

The evolution equations for $\delta d,\,\delta E$ and $\delta \Omega_2$ at linear order are obtained
by perturbing (\ref{appdb})-(\ref{appdomega}):
\begin{align}
&\frac{d}{du} \delta d=\frac{2({\bar E}-1)}{3u}\delta d+\frac{2 {\bar d}}{3u}\delta E, \\
&\frac{d}{du} \delta E= \frac{1}{u} \bigg\{ 1-(4-{\bar \Omega_2})\frac{\bar E}{3} \bigg\} \delta E+\frac{\bar E}{u} \bigg\{ -\frac{4-{\bar \Omega_2}}{3} \delta E+\frac{\bar \Omega_2}{3} \delta \Omega_2 \bigg\}, \\
&\frac{d}{du} \delta \Omega_2= \bigg\{ (1-{\bar \Omega_2}) \frac{2 {\bar E}}{3u}-1 \bigg\}+{\bar \Omega_2} \bigg\{ \frac{2(1-{\bar \Omega_2})}{3u} \delta E-\frac{2{\bar E}}{3u} \delta \Omega_2 \bigg\}.
\end{align}
Behaviors of $\delta d,\,\delta E$ and $\delta \Omega_2$ for $u \ll 1$ are obtained as follows.
Combining (\ref{appscalef2}) and (\ref{appdefd}) yields the time evolution of $\delta d(s)$ for $u \ll 1$ as
\begin{equation}
\delta d(u) =\frac{\Gamma_2^{2/3}}{ {\left( \frac{9}{4} \Omega_2 (t_c) H^2(t_c) \right)}^{1/3} \Omega_2 (t_c) u^{2/3}}.
\end{equation}
For $\delta E$, by using a relation $\delta E = \frac{3t}{2} \delta H$, we find that
\begin{equation}
\delta E (u)=-\frac{\Gamma_2^{2/3}}{ {\left( \frac{9}{4} \Omega_2 (t_c) H^2(t_c) \right)}^{1/3} \Omega_2 (t_c) u^{2/3}}.
\end{equation}
For $\delta \Omega_2$, since it is sourced by the residual radiation, we have
\begin{equation}
\delta \Omega_2(u) = \frac{\rho_r (t)}{\rho_1 (t)}=\frac{a(t_c)}{a(t_d) \Omega_2 (t_c)}=-\frac{\Gamma_2^{2/3}}{ {\left( \frac{9}{4} \Omega_2 (t_c) H^2(t_c) \right)}^{1/3} \Omega_2 (t_c) u^{2/3}}.
\end{equation}
To simplify the system further, let us introduce the new variables by
\begin{align}
X &\equiv {\left( \frac{9}{4} \Omega_2 (t_c) H^2(t_c) \right)}^{1/3} \frac{\Omega_2 (t_c)}{\Gamma_2^{2/3}} u^{2/3} \delta d, \\
Y &\equiv {\left( \frac{9}{4} \Omega_2 (t_c) H^2(t_c) \right)}^{1/3} \frac{\Omega_2 (t_c)}{\Gamma_2^{2/3}} u^{2/3} \delta E, \\
Z &\equiv {\left( \frac{9}{4} \Omega_2 (t_c) H^2(t_c) \right)}^{1/3} \frac{\Omega_2 (t_c)}{\Gamma_2^{2/3}} u^{2/3} \delta \Omega_2.
\end{align}
Then the evolution equations for these variables become
\begin{align}
\frac{dX}{du}&=\frac{2}{3u} \left( {\bar E}X+{\bar d}Y \right), \\
\frac{dY}{du}&=\frac{2}{3u}Y+\frac{1}{u} \bigg\{ 1-\frac{\bar E}{3} (4-{\bar \Omega_2}) \bigg\} Y+\frac{\bar E}{3u} \left( -(4-{\bar \Omega_2})Y+{\bar E}Z \right),\\
\frac{dZ}{du}&=\frac{2}{3u}Z+\bigg\{ \frac{2 {\bar E}}{3u} (1-{\bar \Omega_2})-1 \bigg\}Z+\frac{2{\bar \Omega_2}}{3u} \left( (1-{\bar \Omega_2})Y-{\bar E}Z \right).
\end{align}
The corresponding initial conditions are $X(0)=-Y(0)=-Z(0)=1$.
It is now clear that the new set of differential equations plus the initial conditions are free of model parameters,
which means that we have extracted the dependence of $F_2$ on the model parameters.
Indeed, we find that $F_2$ can be written as
\begin{equation}
F_2 = a_* {\left( \frac{9}{4} \Omega_{2,*} H_*^2 \right)}^{1/3} \frac{c_\Gamma}{\Gamma_2^{5/3}}+a_* {\left( \frac{9}{4} \Omega_{1,*} H_*^2 \right)}^{1/3} \frac{9d_\Gamma^4 f_\Gamma}{16} \frac{\Omega_{1,*}}{\Gamma_1^{2/3} \Gamma_2 \Omega_{2,*}}, \label{appFb}
\end{equation}
where $f_\Gamma$ is a constant defined by
\begin{equation}
f_\Gamma \equiv \int_0^\infty du\, X(u) e^{-u} \approx 1.18716.
\end{equation}
The second term in (\ref{appFb}) comes from $\delta a$.
Substituting (\ref{appFa}) and (\ref{appFb}) into (\ref{app1}), we get
\begin{equation}
N=\frac{1}{4} \log \bigg\{ \left( 1+\epsilon_\Gamma \right) {\left( \frac{H_*}{\Gamma_1} \right)}^{2/3} \Omega_{1,*}^{4/3} +{\left( \frac{H_*}{\Gamma_2} \right)}^{2/3} \Omega_{2,*}^{4/3} \bigg\} +\frac{1}{4}\log \bigg\{ {\left( \frac{9}{4} \right)}^{1/3} a_* c_\Gamma \bigg\} +\frac{1}{4} \log \frac{\rho_*}{\rho_f},
\end{equation}
where $\epsilon_\Gamma$ is a numerical constant defined by
\begin{equation}
\epsilon_\Gamma \equiv \frac{9 d_\Gamma^4 f_\Gamma}{16 c_\Gamma}.
\end{equation}
Numerically, we find that $\epsilon_\Gamma \approx 1.183$.
Since the last two terms are merely constants, only the first term contributes to the curvature perturbation.

\end{document}